\documentclass[aps,prx,superscriptaddress,twocolumn]{revtex4}
\usepackage{graphics}
\usepackage{epsfig}
\usepackage{float}
\usepackage{color}
\usepackage{stackrel}
\usepackage{amsfonts}
\usepackage{wrapfig}
\usepackage{pstricks}
\usepackage{pst-node}
\usepackage{bm}
\usepackage{mathtools}
\usepackage{dcolumn}
\usepackage{multirow}
\usepackage{epstopdf}
\usepackage{subfigure}
\usepackage{amssymb}
\usepackage{amsmath}
\usepackage{commath}
\usepackage{graphicx,bm}
\usepackage{verbatim}

\begin{document}
\title{Marrying excitons and plasmons in monolayer transition-metal dichalcogenides}

\author{Dinh Van Tuan}
\altaffiliation{vdinh@ur.rochester.edu}
\affiliation{Department of Electrical and Computer Engineering, University of Rochester, Rochester, New York 14627, USA}

\author{Benedikt Scharf}
\affiliation{Department of Physics, University at Buffalo, State University of New York, Buffalo, NY 14260, USA}
\affiliation{Institute for Theoretical Physics, University of Regensburg, 93040 Regensburg, Germany}
\affiliation{Institute for Theoretical Physics and Astrophysics, University of W\"{u}rzburg, Am Hubland, 97074 W\"{u}rzburg, Germany}

\author{Igor \v{Z}uti\'c}
\affiliation{Department of Physics, University at Buffalo, State University of New York, Buffalo, NY 14260, USA}

\author{Hanan~Dery}
\altaffiliation{hanan.dery@rochester.edu}
\affiliation{Department of Electrical and Computer Engineering, University of Rochester, Rochester, New York 14627, USA}
\affiliation{Department of Physics and Astronomy, University of Rochester, Rochester, New York 14627, USA}

\begin{abstract} 
Just as photons are the quanta of light, plasmons are the quanta of orchestrated charge-density oscillations in conducting media. Plasmon phenomena in normal metals, superconductors and doped semiconductors are often driven by long-wavelength Coulomb interactions. However, in crystals whose Fermi surface is comprised of disconnected pockets in the Brillouin zone, collective electron excitations can also attain a shortwave component when electrons transition between these pockets.  In this work, we show that the band structure of  monolayer transition-metal dichalcogenides gives rise to an intriguing mechanism through which shortwave plasmons are paired up with excitons.  The coupling elucidates the origin for the optical side band that is observed repeatedly in monolayers of WSe$_2$ and WS$_2$ but not understood.  The theory makes it clear why exciton-plasmon coupling has the right conditions to manifest itself distinctly only in the optical spectra of electron-doped tungsten-based monolayers.
\end{abstract}
\maketitle

\section{Introduction and motivation} \label{sec:intro}

Transition-metal dichalcogenides become direct band-gap semiconductors when thinned to a single atomic monolayer \cite{Mak_PRL10,Splendiani_NanoLett10}. Stacking such two-dimensional crystals via van der Waals forces alleviates the need to fabricate logic and optoelectronic devices with lattice-matched crystals \cite{Wang_NatNano12,Geim_Nature13,Mak_NatPhoton16}.  Furthermore, the lack of space inversion symmetry in these monolayers lifts the spin degeneracy at the edges of the conduction and valence bands \cite{Xiao_PRL12,Song_PRL13}. As a result, time reversal partners the spin and valley degrees of freedom such that spin-up charge carriers populate valleys (low-energy pockets) on one side of the Brillouin zone, while spin-down ones populate valleys on the opposite side \cite{Xiao_PRL12}. The intense recent  research on the spin-valley partnership has led to understanding of optical selection rules and transport phenomena in monolayers transition-metal dichalcogenides (ML-TMDs) \cite{Mak_Science14,Xu_NatPhys14}.   

\begin{figure}
\centering
\includegraphics*[width=9cm]{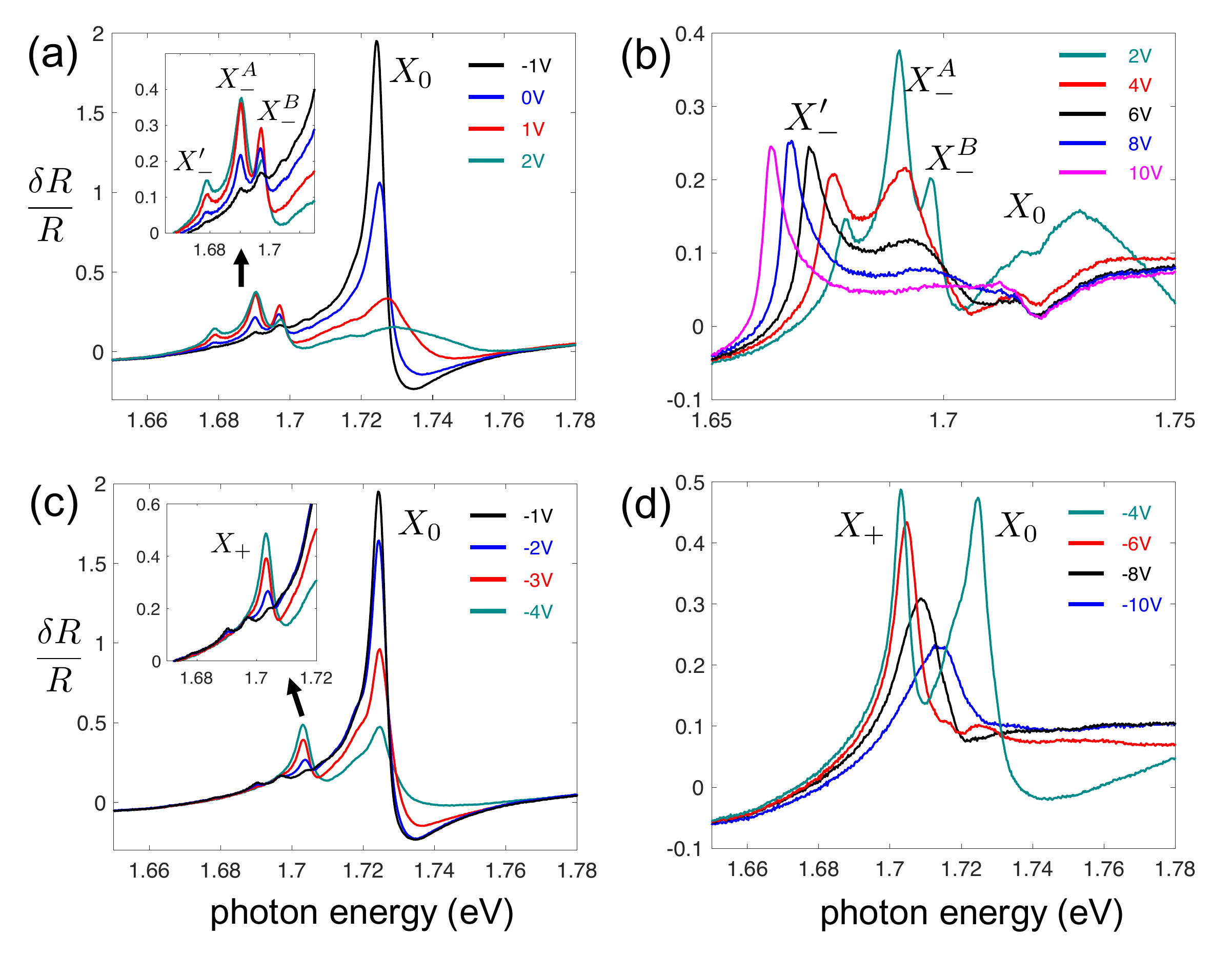}
\caption{Optical reflection spectroscopy of ML-WSe$_2$ at 4~K. The gate voltage controls the charge density in the ML. (a) and (b) Optical spectra for low and high electron densities, respectively. The optical side band is indicated by $X-'$. (c) and (d) The respective results for hole densities. The sample is $n$-type, so that a gate voltage of $\sim$$\,$$-$1~V is required to deplete the ML from resident electrons. These results were kindly provided to us by Zefang Wang, Kin Fai Mak, and Jie Shan. Ref.~\cite{Wang_NanoLett17} includes details of the experiment with a similar device.  }
\label{fig:exp}
\end{figure}

In this work, we focus on a unique optical side band that emerges in electron-doped ML-WSe$_2$ and ML-WS$_2$ \cite{Jones_NatNano13,Xu_NatPhys14,Shang_ACSNano15,Plechinger_PSS15,Plechinger_NanoLett16,Plechinger_NatCom16,Wang_NanoLett17,Wang_NatNano17,Koperski_nanoph17}. Understanding its microscopic origin  is an outstanding open question ever since its  first observation  in the  photoluminescence (PL) measurements of ML-WSe$_2$ by Jones \textit{et al.} \cite{Jones_NatNano13}. 
It was also observed in the PL of electron-doped ML-WS$_2$, and was originally attributed to biexcitons or defects \cite{Shang_ACSNano15,Plechinger_PSS15}. To explain the behavior of this optical side band, we use the empirical data of differential reflectance spectroscopy of ML-WSe$_2$ at 4~K, as shown in Fig.~\ref{fig:exp} where this band is indicated by $X-'$. The monolayer was embedded between thin layers of hexagonal boron-nitride in a dual-gated structure \cite{Wang_NanoLett17}. The bias voltage was the same in the top and bottom gate electrodes, so that only the charge density in the monolayer is controlled while no out-of-plane electrical field is introduced. The relation between gate voltage and charge density is such that 1~V corresponds to $\sim$10$^{12}$~cm$^{-2}$ \cite{Wang_NanoLett17}.

Figure~\ref{fig:exp}(a) shows the spectrum for low and intermediate electron densities. The neutral-exciton spectral line, denoted by $X_0$, dominates the spectrum and its intensity decays with increasing the electron density. Also noticeable are three smaller spectral lines in the low-energy side of the spectrum (as highlighted in the enlarged box). These features cannot be completely quenched due to disordered regions that prevent complete charge depletion across the entire sample \cite{Efros_SSC89}. The spectral lines that are denoted by $X^A_-$ and  $X^B_-$ were recently attributed to two types of negatively charged excitons in tungsten-based monolayers \cite{Jones_NatPhys16,Plechinger_NatCom16,Courtade_arXiv17}. The lowest-energy spectral line, denoted by $X-'$, is the optical side band. In addition to the fact that $X^A_-$ and $X^B_-$ are clearly resolved from that of $X-'$ even at low densities, as highlighted in the enlarged box of Fig.~\ref{fig:exp}(a),  their behavior is strikingly different, rendering their identification relatively easy. For example, $X^A_-$ has the known properties of trions in semiconductor quantum wells \cite{Bar-Joseph_SST05,Suris_PSSB01,Ramon_PRB03,Kossacki_JPCM03,Portella-Oberli_PRB04,Keller_PRB05}: It is amplified without enhanced broadening when the electron population increases in the low-density regime. That is, in the regime that the Fermi energy is measurably smaller than the binding energy of the charged exciton (energy difference between $X^A_-$ and $X_0$, which is about 30-35~meV). Figures~\ref{fig:exp}(b) shows that $X^A_-$ starts to blueshift, decay and broaden at intermediate and elevated densities, in which the Fermi energy is no longer negligible compared with the charged-exciton binding energy. In this density regime, the trion cannot avoid many-body interactions with the Fermi sea. The optical side band, on the other hand, shares none of these features. Figure~\ref{fig:exp}(b) shows that when the electron density continues to increase, $X-'$ redshifts and possibly drains part of the oscillator strength of the decaying charged exciton spectral lines. Its spectral line shows no evident broadening, and its saturated intensity at elevated densities hardly decays (whereas the other spectral lines strongly decay, blueshift and broaden).  

Figures~\ref{fig:exp}(c) and (d) show the absorption spectrum in hole-doped conditions. It includes the neutral and charged-exciton spectral lines ($X_0$ and $X_+$), where the latter behaves similarly to $X^A_-$ in terms of its decay, blueshift and broadening at elevated densities.  There is no optical side band. The reflectivity spectra of ML-MoSe$_2$ is similar (not shown) \cite{Wang_NanoLett17}. It only includes two spectral lines, both in the electron and hole doping regimes, which  behave similarly to $X_0$ and $X_+$ in Figs.~\ref{fig:exp}(c) and (d). They share no common properties with the optical side band of tungsten-based MLs. It is emphasized that in this work we neither model nor focus on the microscopic origin of the charged exciton. Sidler {et al.} recently suggested that it originates from Fermi polarons wherein the exciton is attracted to the surrounding Fermi sea via weak van der Waals forces \cite{Sidler_NatPhys17}. Jadczak {et al.}, on the other hand, attributed the charged exciton to trions, and further elucidated its strong coupling to phonons due to the resonance between the trion binding energy and that of the homopolar phonon mode \cite{Jadczak_PRB17}. 

The observed behavior of the optical side band in tungsten-based compounds is largely identical in reflectance and PL experiments, including the unique feature that the band redshifts in both cases \cite{Jones_NatNano13,Shang_ACSNano15,Wang_NanoLett17}. One noticeable difference, however, is the relative amplitude of the signal. Specifically, reflectance measurements show that the largest oscillator strength is that of the neutral exciton at low densities \cite{Chernikov_PRL15,Cadiz_PRX17,Courtade_arXiv17,Wang_NatNano17,Wang_NanoLett17}. The oscillator strength of the optical side band, on the other hand, is evidently weaker as one can notice by comparing the $y$-axis scales in Figs.~\ref{fig:exp}(a) and (b). The scenario is opposite in PL measurements, where one finds that the strongest  light emission comes from the optical side band at elevated electron densities, while that of the neutral exciton at low densities is evidently weaker  \cite{Jones_NatNano13,Shang_ACSNano15,Wang_NanoLett17}. This apparent contradiction is settled by the facts that in PL experiments: (i) Neutral excitons in tungsten-based monolayers experience intervalley scattering after photoexcitation, rendering them less optically active in the light emission process  \cite{Dery_PRB15,Zhang_PRL15,Wu_PRB15,Dery_PRB16}. (ii) The quantum efficiency of light emission can be improved at elevated electron densities due to screening of charged defects that function as non-radiative recombination centers for excitons. (iii) We will show that several photon emission mechanisms contribute to the optical side band in the PL case.

The main contribution of this work is in elucidating the microscopic origin for the optical side band, revealing an intriguing pairing phenomenon between excitons and plasmons, which has no  parallel in other known two-dimensional (2D) systems.  We make it clear why the optical band is manifested in the spectra of electron-doped ML-WSe$_2$ and ML-WS$_2$, while being absent in the hole-doped case or in ML-MoSe$_2$ and ML-MS$_2$. The theoretical framework in this work studies the oscillator strength of excitons through the electron-hole pair function  in the presence of dynamical screening. The theory extends beyond the quasi-static random-phase approximation or the Shindo approximation  for the dynamical effects  \cite{Haug_SchmittRink_PQE84}. It therefore allows one to model the decay of the neutral exciton at elevated charge densities without introducing phenomenological  screening parameters or ansatzes regarding the energy dependence of the pair function. In addition, the theory allows one to model the emergence of dynamical optical bands, where we will focus on the one that arises due to the exciton-plasmon coupling at large electron densities.


\section{Basic properties of Excitons and Plasmons in ML-TMDs}\label{sec:basics}

\begin{figure}
\centering
\includegraphics*[width=8cm]{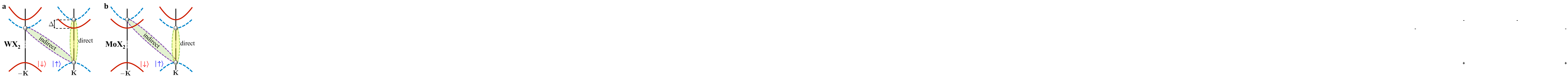}
\caption{Direct and indirect excitons in the time-reversed $K$-point valleys. (a) and (b) show the involved electronic states in tungsten- and molybdenum-based monolayers, respectively. The indirect exciton has lower energy in the former case. Spin of the bands is color coded, and $\Delta$ is the conduction-band splitting energy. }
\label{fig:excitons}
\end{figure}

Before we can explain the coupling between plasmons and excitons, we briefly summarize the relevant  properties of each of these quasiparticles separately.  
Excitons in ML-TMDs are classified based on the spin and valley configuration of the electron-hole pair \cite{Dery_PRB15,Wu_PRB15,Zhang_PRL15}. Figure~\ref{fig:excitons} shows the valleys pertinent to the formation of low-energy excitons in ML-WX$_2$ and ML-MoX$_2$ where X is the chalcogen atom.  
Excitons are bright or dark (optically active or inactive) if the spins of the electron in the conduction band and the missing electron in the valence band are parallel or antiparallel, respectively. Scattering between dark and bright excitons necessitates a spin-flip of the electron or hole, which is typically a much slower process than the lifetime of bright excitons \cite{Lagarde_PRL14,Wang_PRB14,Yang_NatPhys15,Song_NanoLett16}. Therefore, we neglect dark excitons and show that the dynamical relation between two types of bright excitons is sufficient to explain the optical measurements in Fig.~\ref{fig:exp}.  The first type is the direct exciton in which the electron and hole have the same valley index, giving rise to direct-gap optical transitions. The second type is the indirect exciton in which the electron and hole reside in opposite valleys.  Optical transitions of indirect excitons in perfect crystals are mediated by external agents such as shortwave phonons, needed to make up for the large momentum mismatch of indirect excitons and photons.  As shown in Fig.~\ref{fig:excitons}, indirect excitons have lower energy than direct ones in ML-WX$_2$ and vice versa in ML-MoX$_2$ \cite{Dery_PRB15}.

\begin{figure}[t]
\centering
\includegraphics[width=7cm]{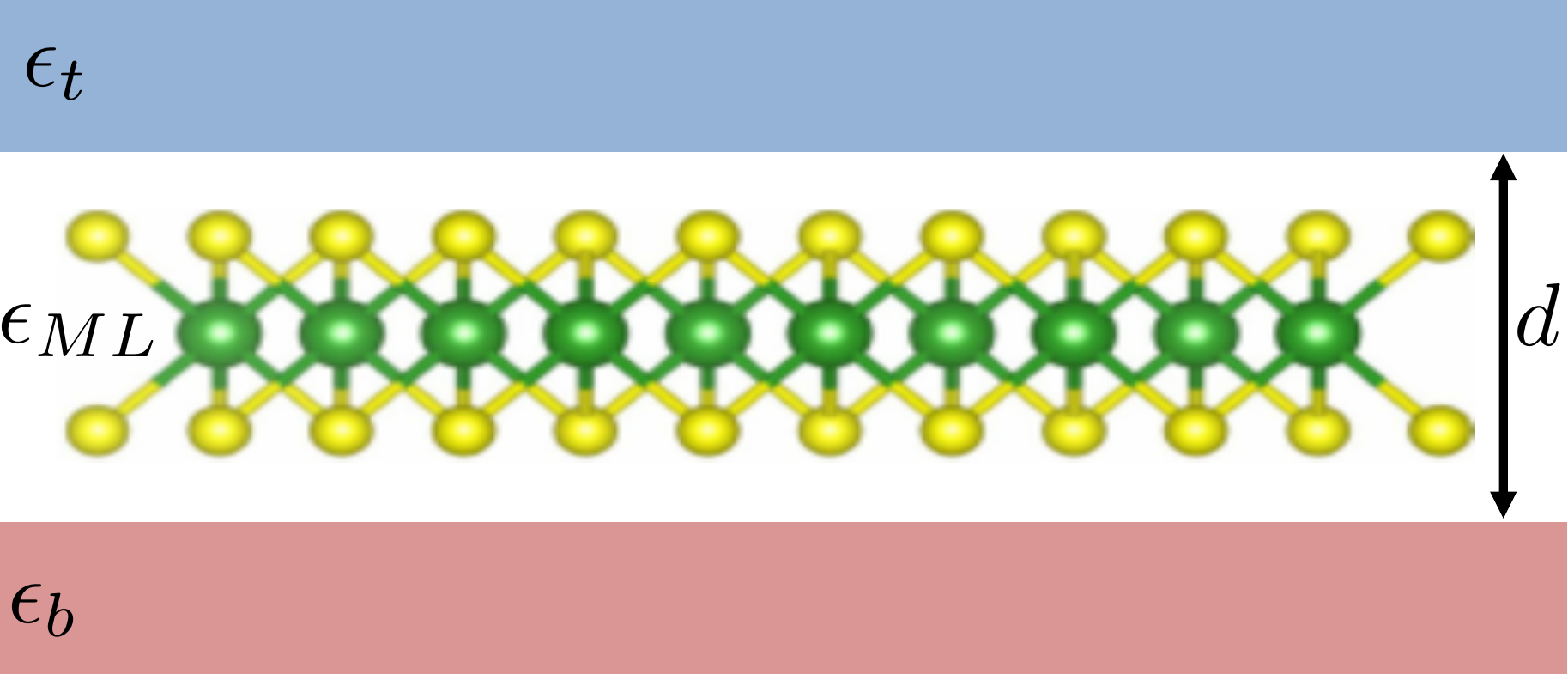}
\caption{ The studied geometry in this work is that of a ML-TMD  sandwiched between  two other materials. The thickness of the ML including the van der Waals gap is $d$. The dielectric constants of the ML, bottom and top layers are $\epsilon_{ML}$, $\epsilon_b$ and $\epsilon_t$, respectively.}\label{fig:Geometry}
\end{figure}

Next, we briefly discuss the two plasmon species in ML-TMDs. The first type of plasmons is characterized by their long-wavelength \cite{Bohm_PR53,Bardasis_PR61,Sun_PRL16}. They originate from the long-range part of the Coulomb potential through which  electrons cross the Fermi surface with small crystal momentum transfer (i.e., an intravalley process). These plasmons are common to all conducting media and are largely independent of the band structure since their wavelength is much longer than the lattice constant \cite{Pines_PR56}. In the context of optical transitions in semiconductors, plasmons screen the electron-hole attraction and they shrink the band-gap energy  by assisting electrons (or holes) of similar spins to avoid each other  \cite{Haug_SchmittRink_PQE84}. The energy dispersion of long-wavelength plasmons in 2D semiconductors follows
\begin{eqnarray}
E_{\ell}(k) = \sqrt{\frac{2e^2E_F k}{\epsilon(k)}}  \,\,\, ,  \label{eq:intra_plasmon_dispersion}
\end{eqnarray}
where $e$ is the elementary charge, $k$ is the wavenumber of the plasmon, and $E_F = \pi \hbar^2 n/m_e$ is the Fermi energy related to the charge density and effective mass of the electron ($n$ and $m_e$). The static dielectric function $\epsilon(k)$ takes into account  the experimental sample geometry, as depicted in Fig.~\ref{fig:Geometry}.  Its form follows \cite{Keldysh_JETP79}
\begin{eqnarray}
\epsilon(k) &=&     \frac{\epsilon_{ML} \left[ \epsilon_{+} \sinh(kd)+\epsilon_{\times}\cosh(kd)\right]}{ \epsilon_{-} + \epsilon_{+}\cosh(kd)+\epsilon_{\times}\sinh(kd)} \,\,,\label{eq:Keldysh_dielectric}
\end{eqnarray}
where $d$ denotes the thickness of the ML including the  van der Waals gap. The dielectric constants follow $\epsilon_{\pm}  = \epsilon_{ML}^2 \pm \epsilon_t \epsilon_b$ and $\epsilon_{\times} = \epsilon_{ML}(\epsilon_t+\epsilon_b)$, where $\epsilon_{ML}$, $\epsilon_b$ and $\epsilon_t$ denote the dielectric constants of the ML, bottom and top layers, respectively. This dielectric function is relevant for short and long wavelengths, where \cite{Cudazzo_PRB11}
\begin{eqnarray}
\epsilon(k) & \xrightarrow{ kd \ll 1}  &  \frac{\epsilon_t+\epsilon_b}{2} +  \left( \epsilon_{ML} - \frac{\epsilon_t^2 + \epsilon_b^2}{2\epsilon_{ML}} \right) \frac{kd}{2} \,,  \,\,\, \nonumber \\
\epsilon(k) & \xrightarrow{ kd \gg 1}  &   \epsilon_{ML}   \,\,\,. 
\end{eqnarray}
 In the long-wavelength regime ($kd \ll 1$), the effective dielectric constant is largely set by the average dielectric constants of the top and bottom layers.  In the opposite extreme, on the other hand, the top and bottom layers play no role since the electric field lines are confined in the ML at such shortwaves.

\begin{figure}
\centering
\includegraphics*[width=8cm]{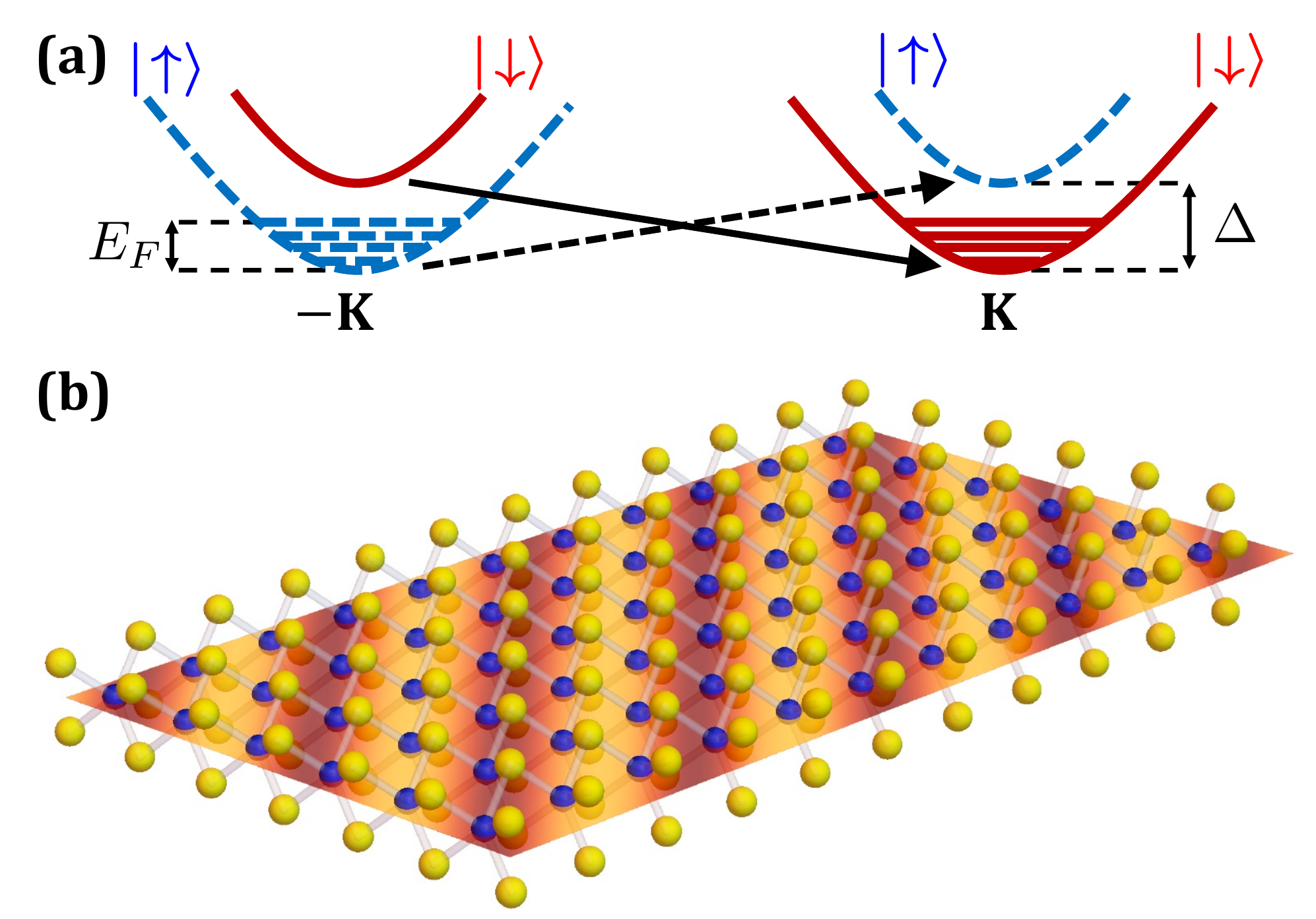}
\caption{The intervalley Coulomb interaction in ML-TMDs. (a) Spin-conserving charge excitations from the $-K$ to the $K$ valleys. Electrons need to overcome the spin-splitting energy gap in the conduction band. (b) The resulting shortwave charge fluctuations in the monolayer.}
\label{fig:plasmons}
\end{figure}

The second type of plasmons originates from the short-range Coulomb potential through which electrons transition between valleys \cite{Adler_PR62,Tudorovskiy_PRB10,Dery_PRB16,Schmidt_NanoLett16,Danovich_SR17}. Due to the relatively large spin splitting in the valence band of ML-TMDs  \cite{Xiao_PRL12}, such transitions are mostly relevant in the conduction band wherein the much smaller spin splitting can be comparable to the Fermi energy, as shown in Fig.~\ref{fig:plasmons}(a). Collective  intervalley excitations result in shortwave plasmons, whose charge fluctuations profile is illustrated in Fig.~\ref{fig:plasmons}(b), and whose energy dispersion follows \cite{Dery_PRB16}
\begin{eqnarray}
E_{\text{s}}(q)   =  \Delta + \left( 1 +   \frac{3}{\alpha} \right)\tilde{\varepsilon}_q  + \frac{\alpha}{3} E_F  \,\,\, .  \label{eq:plasmon_dispersion}
\end{eqnarray}
$\Delta$ is the spin-splitting energy in the conduction band, $\alpha = (K_0 a_{\text{s}})^{-1} \sim 0.1$ , and $\tilde{\varepsilon}_q = \hbar^2 (K_0 - q)^2 / 2m_e$. Here, $K_0 = 4\pi/3a$ is the wavenumber that connects the time-reversed valleys ($a\,$$\sim\,$3.2~$\AA$ is the lattice constant), and  $a_{\text{s}} = \hbar^2 \epsilon_{ML} /e^2m_e$ is the effective Bohr radius at these short wavelengths. The region of free-plasmon propagation is limited to $K_0 - \tfrac{\alpha}{3} k_F  < q < K_0 +  \tfrac{\alpha}{3} k_F$ where $k_F$ is the Fermi wavenumber \cite{Dery_PRB16}. That is, when the  wavenumber of a shortwave plasmon belongs to a small region of nearly perfect time-reversal intervalley conditions ($q \sim K_0 $), it does not interact with individual electrons and should be treated as an independent degree of freedom in the electronic system \cite{Pines_PR56}.

In addition to their different length scales, intravalley and intervalley plasmons in ML-TMDs differ in their energy spectrum. The spin splitting  in the conduction band produces a gapped dispersion for shortwave plasmons compared with the gapless dispersion in the long wavelength case:  $E_{\text{s}}(K_0) = \Delta + \alpha E_F/3$ while $E_{\ell}(0)=0$. The value of $\Delta$ is governed by the spin-orbit coupling at vanishing electron densities.  Its value increases at elevated densities due to the electron-electron exchange interaction \cite{Dery_PRB16}.

\begin{figure*}[t]
\includegraphics[width=15cm]{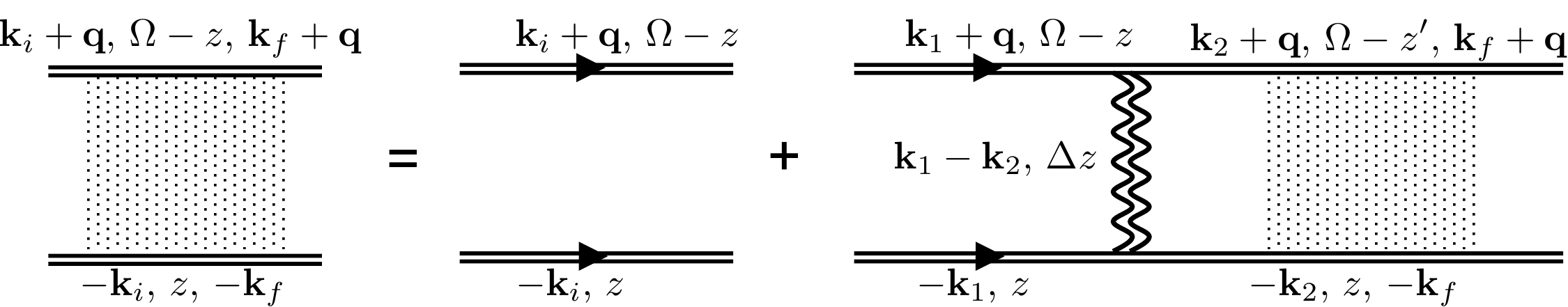}
\caption{Feynman diagram representation of the two-particle Dyson equation under the screened ladder approximation. The top and bottom propagators denote the Green's functions of electrons and holes, respectively. The wiggly line denotes the dynamically-screened attractive potential.  This diagram sums over all processes through which the electron-hole pair is repeatedly scattered by the attractive potential.   }\label{fig:BSE_Dia}
\end{figure*}

\section{The Pair function of Excitons in the presence of Plasmons}

To study how excitons interact with plasmons, we use the finite-temperature Green's functions formalism \cite{Mahan_Book}, and quantify the optical signature of bright excitons when surrounded by collective charge excitations \cite{Haug_SchmittRink_PQE84}. The connection to the absorption spectrum is then established via the imaginary part of the electron-hole pair function, 
\begin{eqnarray}
A(\omega) \propto  \beta^{-1} \sum_{\mathbf{k},z} \text{Im} \left\{G_p(\mathbf{q} \rightarrow 0, \mathbf{k}, z, \Omega \rightarrow  E_{\omega} + i \delta) \right\}. \label{eq:abs}
\end{eqnarray}
$\hbar \omega$ is the photon energy and  $\beta^{-1} = k_BT$ is the thermal energy.   $\mathbf{k}$ and $\mathbf{q}$ are 2D crystal wavevectors,  where the electron in the pair takes on $\mathbf{k} +\mathbf{q}$  and the hole  -$\mathbf{k}$. In the limit that $\mathbf{q} \rightarrow 0$, a direct optical transition is rendered due to the small momentum of photons.   Throughout the analysis below, pair functions of direct and indirect excitons are evaluated by assigning $q=0$ and $q=K_0$, respectively.  Dynamical effects due to plasma oscillations are embodied by the dependence of $G_p$ on $z$ and $\Omega$, which are odd (Fermion) and even (Boson) imaginary Matsubara energies, $z=(2\ell+1)\pi i/\beta$ and $\Omega=2\ell \pi i/\beta$, where $\ell$ is an integer \cite{Haug_SchmittRink_PQE84}. The sums over $\mathbf{k}$ and $z$ in Eq.~(\ref{eq:abs}) integrate out the Fermion degrees of freedom. In the last step, Pad\'{e} analytical continuation is performed in order to evaluate the pair function at real photon energies \cite{Vidberg_JLTP77}:  $ \Omega \rightarrow  E_{\omega}  + i \delta$ where $\delta \rightarrow 0^{+}$ and  $E_{\omega} = \hbar \omega - \mu_e -\mu_h$ is expressed in terms of photon energy and quasi-chemical potentials in the conduction and valence bands. 

To find the pair function in Eq.~(\ref{eq:abs}), we make use of the so-called screened ladder approximation. It describes repeated interaction of the electron-hole pair with the dynamically-screened potential \cite{Haug_SchmittRink_PQE84}. While this physical picture considers the interaction of excitons with collective electron excitations, it neglects exciton-exciton interactions. Accordingly, the screened ladder approximation can be used to explain experiments in which the ML is not subjected to intense photoexcitation. Figure~\ref{fig:BSE_Dia} shows the Feynman diagram of the two-particle Green function before contraction (i.e., before integrating out any of its quantum numbers). This diagram is formally written as
\begin{widetext}
\begin{eqnarray}
 G_{p}(\mathbf{q}, \mathbf{k}_i, \mathbf{k}_f,z,\Omega)= G_{p,0}(\mathbf{q}, \mathbf{k}_i, \mathbf{k}_f, z, \Omega)+\frac{1}{\beta}\sum_{\mathbf{k}_1,  \mathbf{k}_2,z'}G_{p,0}(\mathbf{q}, \mathbf{k}_i,  \mathbf{k}_1 , z, \Omega) V_s( \mathbf{k}_1-  \mathbf{k}_2, z - z' )G_{p}( \mathbf{q},  \mathbf{k}_2, \mathbf{k}_f,z',\Omega)  \,,  \label{BSEEq1}
\end{eqnarray}
\end{widetext}
where $V_s({\bf  k-k}',z-z')$ is the dynamically-screened potential to be discussed later. $G_{p}(\mathbf{q}, \mathbf{k}_i, \mathbf{k}_f,z,\Omega)$ and  $G_{p,0}(\mathbf{q}, \mathbf{k}_i, \mathbf{k}_f, z,\Omega)$ are the interacting and non-interacting pair functions, respectively. The non-interacting case is given by the product of the electron and hole Green's function 
\begin{equation}
G_{p,0}(\mathbf{q}, \! \mathbf{k}_i, \! \mathbf{k}_f, \!z, \!\Omega)\!=\!G_{e}(\mathbf{k}_i\!+\!\mathbf{q},\Omega\!-\!z)G_{h}(-\mathbf{k}_i, \!z )\delta_{\mathbf{k}_i, \mathbf{k}_f},
\end{equation} 
where here and below we assume the area of the monolayer  is 1. The Kronecker delta-function allows us to integrate out the final wavevector $\mathbf{k}_f$ in Eq.~(\ref{BSEEq1}), and calculate the contracted pair function, $G_{p}(\mathbf{q}, \mathbf{k},z,\Omega)$, from which we can estimate the absorption according to Eq.~(\ref{eq:abs}). We note that while this step seems trivial, the approach so-far was to contract over the odd-Matsubara energy ($z$) and then use the so-called Shindo approximation to describe  dynamical screening \cite{Shindo_JPSJ70,Haug_SchmittRink_PQE84}. However, dynamical screening can be better described upon contraction in momentum rather than energy space. This contraction keeps both dynamical parameters  (even and odd Matsubara energies), while eliminating  $\mathbf{k}_f$ does not compromise any further accuracy in the solution of the pair function if the starting point is the screened ladder approximation. After contraction, we simplify the form of Eq.~(\ref{BSEEq1}),
\begin{eqnarray}
&& \!\!\!\! \!\! \!\! \!\!  G_p(\mathbf{q}, \mathbf{k}, z, \Omega)  = G_e( \mathbf{k}+\mathbf{q}, \Omega - z)G_h( -\mathbf{k}, z)  \times  \nonumber \\ 
&& \,\, \left\{   1 + \beta^{-1} \sum_{\mathbf{k}',z'}   V_s(\mathbf{k}-\mathbf{k}', z-z') G_p(\mathbf{q}, \mathbf{k}', z', \Omega)   \right\}\!\!. \,\,\,\,\,\,  \label{eq:pair_Dyson}
\end{eqnarray}
This equation can be solved by matrix inversion techniques, as explained in Appendix~\ref{Sec:Computational}.

\subsection{The effect of long wavelength plasmons on excitons}\label{sec:only_intra}

We begin by evaluating the pair functions of direct excitons when $V_s$, $G_e$ and $G_h$ in Eq.~(\ref{eq:pair_Dyson}) are affected only by long-wavelength plasmons. Neglecting shortwave plasmons in the first step allows us to find common attributes between ML-TMDs and conventional semiconductor quantum wells \cite{HaugKoch_Book}. 

A central part of the theory is the inclusion of the dynamically-screened potential in Eq.~(\ref{eq:pair_Dyson}).  We invoke the single-plasmon pole approximation (SPP), which provides a relative compact form for the dynamical screening. Using the more rigorous random-phase approximation does not lead to qualitative changes or to increased computational complexity when solving Eq.~(\ref{eq:pair_Dyson}).  When including long-wavelength plasma excitations, the dynamically-screened Coulomb potential under the SPP approximation reads \cite{Haug_SchmittRink_PQE84}
\begin{equation}
 V_s({\bf k},z-z')=\frac{2\pi e^2}{\epsilon(k)}\frac{1}{k}\left( 1+\frac{ E_{\ell}^2(k) }{(z-z')^2- E_k^2} \right) \,, \label{eq:SPP}
\end{equation}
where
\begin{equation}
 E_k =  E_{\ell}(k) \sqrt{ 1 + \frac{a_{\ell} k}{4}} \,, \label{eq:pole}
\end{equation}
is the pole energy, and  $a_{\ell} = \hbar^2(\epsilon_{b}+\epsilon_{t})/2e^2m_e$ is the effective Bohr radius at long wavelengths. The solution of Eq.~(\ref{eq:pair_Dyson}) is greatly simplified when using the static approximation, $V_s({\bf k},0)$, or when studying the zero density case, $E_{\ell}(k) = 0$. In these cases, one first calculates the sum $\sum_z G_e( \mathbf{k}+\mathbf{q}, \Omega - z)G_h( -\mathbf{k}, z)$, and then directly calculates the absorption spectrum from the contracted pair function, $G_p(\mathbf{q}, \mathbf{k}, \Omega)$. The static approximation, however, tends to overestimate the role of screening, and one has to introduce phenomenological terms in the static potential to mitigate the screening effect \cite{SchmittRink_PRB86,HaugKoch_Book}. A better description for the screening is provided by keeping the dynamical terms in the potential, $V_s({\bf k},z-z')$. This approach, however,  renders the solution of Eq.~(\ref{eq:pair_Dyson}) evidently more computationally demanding compared with the static case.

The second signature of  long-wavelength plasmons is introducing the band-gap renormalization (BGR) through the self-energy terms in the electron and hole Green's functions, 
\begin{equation}
G_{i}({\bf k},z)=\frac{1}{z-\varepsilon_{{\bf k},i}-\Sigma_{i}({\bf k}, z)+\mu_{i}},
\end{equation}
where $i=\{e,h\}$, and $\varepsilon_{{\bf k},i} = \hbar^2 k^2/2m_i$ is the noninteracting electron (or hole) energy. The self energy term, $\Sigma_{i}({\bf k}, z)$, stems from the interaction of the electron (or hole) with the dynamically-screened potential, and $\mu_i = \mu_{i,0} + \Sigma_{i}({\bf k}_F, E_F-\mu_{i,0})$ is the chemical potential. $\mu_{i,0}$ is the temperature and density dependent chemical potential of the non-interacting electron gas. Using the $GW$-approximation, the self-energy term is written as \cite{Haug_SchmittRink_PQE84}  
\begin{eqnarray}
\Sigma_{i}({\bf k}, z)&=&-\beta{^{-1}} \sum_{{\bf{q}},\Omega}  V_s({\bf q},\Omega)G_{i,0}({\bf k-q},z-\Omega) \nonumber \\
&=& \varSigma_{x,i}({\bf k})+\varSigma_{c,i}({\bf k},z), \label{eq:self_general}
\end{eqnarray}
where $G_{i,0}$ is the Green's function of the noninteracting particle ($\Sigma = 0$).  The self energy is separated into exchange and correlation parts. The former follows
\begin{equation}
  \varSigma_{x,i}({\bf k})=-\sum_{{\bf q}} \frac{2\pi e^2}{\epsilon(q)}\frac{f_i({\bf k-q})}{q}   \,,
\end{equation}
where $f_i$ is the Fermi-Dirac distribution.  The exchange contribution from electron-electron interaction can only shift the filled valleys (i.e., $f_i \neq 0$). The correlation term, on the other hand, affects all bands and it comes from emission or absorption of long-wavelength plasmons (intravalley processes),
\begin{eqnarray}
  \varSigma_{c,i}({\bf k},z) &=& \sum_{{\bf q} < q_c} \frac{2\pi e^2}{\epsilon(q)}\frac{1}{q} \frac{E_{\ell}^2(q)}{2E_q}  \times   \label{eq:correlation} \\ &&\left\{ \frac{f_i({\bf k-q})+g(E_q)}{z-\varepsilon_{{\bf k-q},i} +\mu_i+E_q} -  (E_q \leftrightarrow -E_q)  \right\} \nonumber  \,,
\end{eqnarray}
where $g(E_q)$ is the Bose Einstein distribution function evaluated at the plasmon pole energy, and $q_c$ is the integration cutoff introduced to guarantee  convergence. Physically, it amounts to avoiding plasma damping due to particle-hole excitations of the Fermi sea.

\begin{figure}
\centering
\includegraphics[width=8cm]{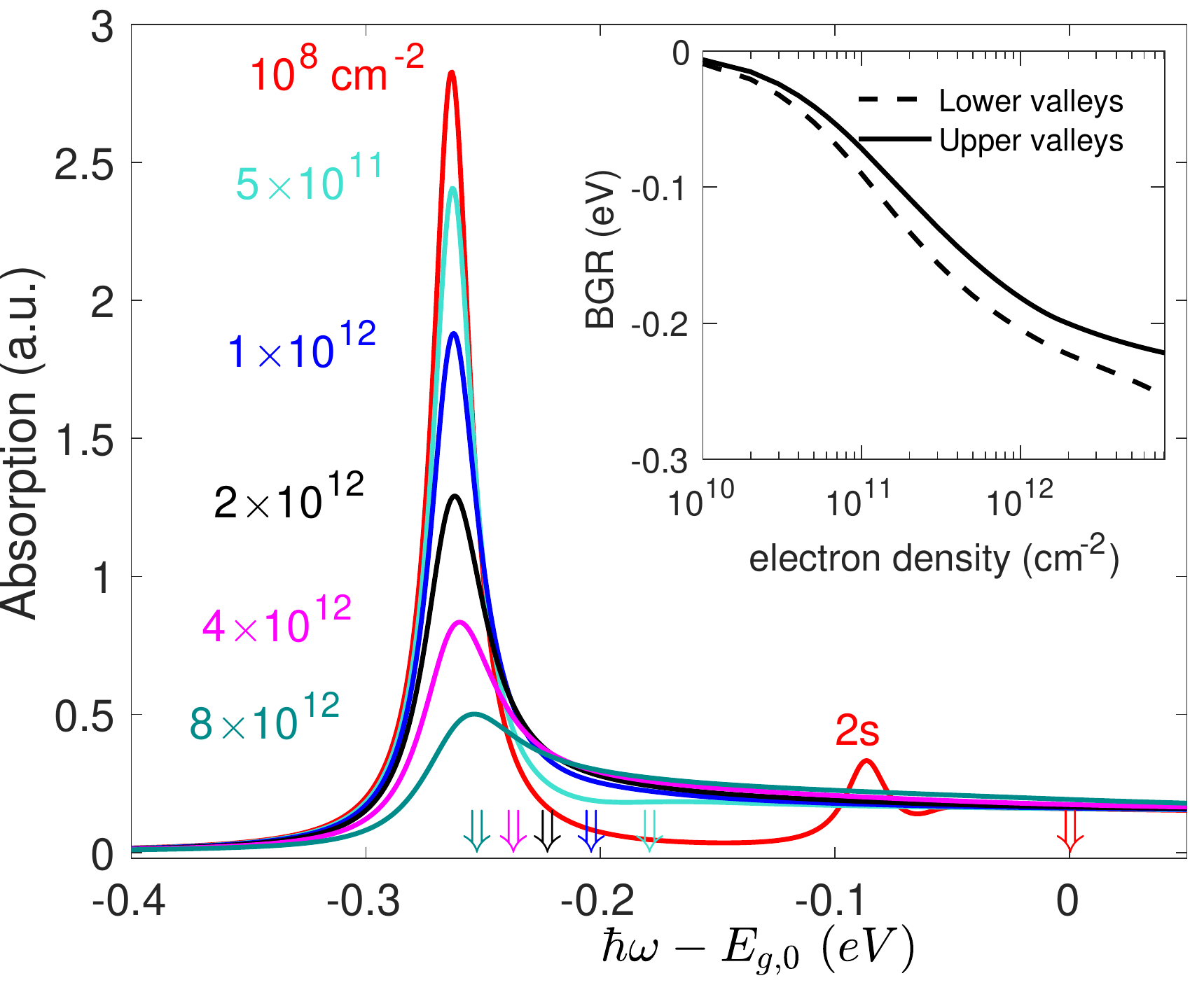}
\caption{ Characteristic effect of long-wavelength plasmons on the absorption spectrum of neutral excitons in ML-TMDs.   Increasing the background charge density leads to a decay of the exciton peak. The  band-gap renormalization (BGR)  is indicated by the arrows on the $x$-axis, showing the redshift for the onset of optical transitions in the continuum (noninteracting electron-hole pairs).  Inset: BGR of the upper and lower valleys as a function of electron density in the lower valleys.  The signatures of dynamical screening and BGR are qualitatively the same when choosing other parameters of effective masses and dielectric constants (Appendix~\ref{Sec:more_results}).}
\label{fig:intra}
\end{figure}

We now have all of the ingredients to solve the pair function in Eq.~(\ref{eq:pair_Dyson}), and then to the calculate the absorption spectrum from Eq.~(\ref{eq:abs}) after performing analytical continuation from the imaginary to the real axis (see Appendix~\ref{Sec:Computational} for more details). Figure~\ref{fig:intra} shows the calculated absorption spectrum at T=10~K from which one can clearly notice the exciton decay when the background charge density increases. The effective masses for both electrons and holes are 0.36$m_0$ \cite{Kormanyos_2DMater15}, where $m_0$ is the free-electron mass (see Appendix~\ref{Sec:mass}). The dielectric environment is modeled by sandwiching the ML ($d=1$~nm, $\epsilon_{ML}=7.25$) between thick layers of hexagonal boron-nitride ($\epsilon_b=\epsilon_t=2.7$) \cite{Wang_NanoLett17,Kim_ACSNano15}. The arrows in Fig.~\ref{fig:intra} indicate the  continuum redshift due to BGR. Above this energy the excitons are no longer bound. Note that the $2s$ bound state merges into the continuum already at relatively small densities due to its smaller binding energy. The BGR is calculated from the value of $\Sigma_{i}({\bf k}, \varepsilon_{{\bf k},i}-\mu_i^0)$ at ${\bf k}=0$, by assuming rigid shift of the bands.  The cutoff energies we have used for the bottom and top spin-split valleys in Eq.~(\ref{eq:correlation}) are  $\varepsilon_{q_c}=60$~meV and $\varepsilon_{q_c}=45$~meV, respectively. These cutoff energies are the same for all densities. Using  larger (smaller) cutoff energies will gradually redshift (blueshift) the peak position as one increases the charge density. The inset of Fig.~\ref{fig:intra} shows that the BGR is strongest when the charge density is ramped up from zero to $\sim\,$10$^{12}$~cm$^{-2}$. The continuum redshift is much slower between $\sim\,$10$^{12}$ and $\sim\,$10$^{13}$~cm$^{-2}$, during which  excitons become unbound and merge into the continuum.


\begin{figure*}[t]
\includegraphics[width=17cm]{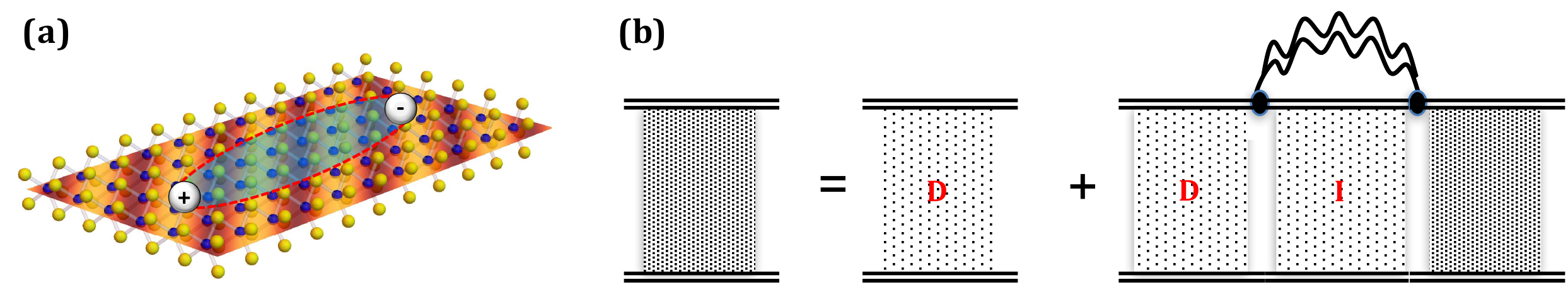}
\caption{ Coupling between excitons and shortwave plasmons. (a) Cartoon of the exciton and shortwave charge fluctuations. The plasmon wavelength is shorter than the exciton extent, which is further blown-up due to screening. (b) Feynman diagram representation of the renormalized direct-exciton pair function due to shortwave plasmons. The self-energy term takes into account the coupling between direct and  indirect excitons (D and I) due to the electron-plasmon interaction. The hole (bottom propagator) acts as a spectator, while the electron interacts with the plasmon. The double wiggly lines in this diagram denote the plasmon propagator and the vertexes denote the electron-plasmon interaction. The diagram is an infinite sum of the virtual processes $D$ + $D\,I\,D$ + $D\,I\,D\,I\,D + ...$. }\label{fig:renorm_Dia}
\end{figure*}

Two effects cause the decay of the neutral-exciton spectral line with the increase in electron density. The dominant one is the broadening effect when the exciton energy approaches the continuum  (see arrows in Fig.~\ref{fig:intra} and Eq.~(\ref{eq:broad}) in the Appendix).  The decay that is caused by reduction in the oscillator strength plays a secondary role due to the mitigated effect of screening in 2D. This behavior is reminiscent of excitonic enhancement of free electron-hole transitions in semiconductor quantum wells \cite{SchmittRink_PRB86}, in which the overlap between the electron and hole wavefunctions remains sizable when excitons enter continuum. The simulated decay of the neutral exciton in Fig.~\ref{fig:intra}  is corroborated in the differential-reflectivity measurements, which resolve the oscillator strength and broadening of the spectral lines \cite{Wang_NanoLett17,Wang_NatNano17}. Figures~\ref{fig:exp}(a) and (c) show that when the charge density is ramped up by gate voltage, the measured decay of neutral excitons is not compensated by equivalent increase of charged excitons, thereby reinforcing the important role of BGR  and dynamical screening. It is noted that unlike the experimental findings of Fig.~\ref{fig:exp}, the simulated exciton peak in Fig.~\ref{fig:intra} is not completely quenched at elevated densities. In addition to the fact that we do not consider charged excitons, which drain part of the oscillator strength of neutral excitons \cite{Esser_PSSB01},  the model tends to overestimate the Coulomb attraction at elevated densities. The reason is that broadening effects are not introduced in the electron and hole Green's functions of Eq.~(\ref{eq:pair_Dyson}); they are introduced in the analytical continuation phase after the pair function is solved (Appendix~\ref{Sec:Computational}).

Additional important result in Fig.~\ref{fig:intra}, which is supported by Figs.~\ref{fig:exp}(a) and (c),   is that the neutral exciton spectral position is hardly affected by the background charge density. The reason is the mutual offset between BGR and the screening-induced reduction in the binding energy of excitons; the long-wavelength plasmons are responsible for both effects \cite{HaugKoch_Book}. One can therefore realize that the binding energy of excitons in ML-TMDs can be of the order of few hundreds meV only at vanishing charge densities. All in all, long-wavelength plasmons lead to qualitatively similar effects in ML-TMDs and conventional semiconductor quantum wells \cite{SchmittRink_PRB86}. The only change is a quantitatively larger effect in ML-TMDs due to reduced dielectric screening and larger electron mass \cite{Ugeda_NatMater14,Chernikov_NatPhoton15}.


\subsection{The effect of shortwave plasmons on excitons}\label{sec:only_intra}

In the next step, the unique features of ML-TMDs are revealed by turning on the coupling between shortwave plasmons and excitons. In the long wavelength case, the exciton-plasmon coupling is implicit:  It is embodied in the self energies of the electron and hole as well as in the attractive screened potential (i.e. the coupling to plasmons is `concealed' in the dressed propagator and Coulomb lines in Fig.~\ref{fig:BSE_Dia}). An explicit exciton-plasmon interaction is inhibited if the extension of the exciton is smaller than the wavelength of the plasmon due to the charge neutrality of the former. This physical picture changes for shortwave plasmons, which can be effectively paired up with an exciton in spite of its charge neutrality. Here, the short-range nature of the charge fluctuations, as can be seen in Fig.~\ref{fig:renorm_Dia}(a), allows the hole or electron to act as a spectator while the plasmon interacts with the opposite charge.

The energy required to excite a shortwave plasmon at low temperatures is greater than that available thermally ($\Delta \gg k_BT$). In addition, we recall that free propagating plasmons are independent degrees of freedom, which do not interact with individual charged particles of the electronic system. As a result, plasmon signatures in the optical spectrum can be resolved by supplying energy from outside the electron system in amounts greater than the plasmon energy. This is the case when we shake-up the electronic system by exciting electron-hole pairs. Figure \ref{fig:renorm_Dia}(b) shows the Feynman diagram for the explicit coupling between an exciton and a shortwave plasmon when the hole acts as the spectator. The plasmon propagator is the double wiggly line. This Feynman diagram corresponds to renormalization of the pair function of direct excitons according to 
\begin{eqnarray}
\tilde{G}_p(\mathbf{q} \rightarrow 0, \mathbf{k},\Omega) = \frac{G_p(\mathbf{q} \rightarrow 0, \mathbf{k},\Omega)}{1-\Sigma_{s}(\mathbf{k},\Omega)G_p(\mathbf{q} \rightarrow 0, \mathbf{k},\Omega)},\,\,\,\, \label{eq:inter_Dyson}
\end{eqnarray}
where the pair Green's functions were contracted by integrating out the Fermion Matsubara energies. The mixing with indirect excitons comes from the self-energy term,
\begin{eqnarray}
\Sigma_{s}(\mathbf{k},\Omega)  =   \beta^{-1}\!\!\sum_{\mathbf{q}',\Omega'}  M_{\mathbf{q}'}^2 D(\Omega - \Omega', \mathbf{q}' )  G_p(\mathbf{q}', \mathbf{k}, \Omega') \,, \label{eq:self_inter_Dyson}
\end{eqnarray}
where $M_{\mathbf{q}'}$ is the interaction between the plasmon and the electron component of the exciton, and $D(\Omega, \mathbf{q} )$ is the free-plasmon propagator. The sum in Eq.~(\ref{eq:self_inter_Dyson}) is restricted to intervalley transitions, $q' \rightarrow K_0$, so that the pair function on the right-hand side is that of indirect excitons. The interaction and propagator terms follow \cite{Pines_PR56}
\begin{eqnarray}
M_{\mathbf{q}'} &=& \frac{\sqrt{3\pi}\hbar^2| \mathbf{K}_0 - \mathbf{q}' |}{m_e}    \,,  \label{eq:electron_plasmon} \\
D(\Omega, \mathbf{q} ) &=& \frac{2E_{\text{s}}(q)}{\Omega^2 - E_{\text{s}}^2(q) }.  \label{eq:plasmon_propagator}
\end{eqnarray}

\begin{figure}[ht!]
\includegraphics[width=9cm]{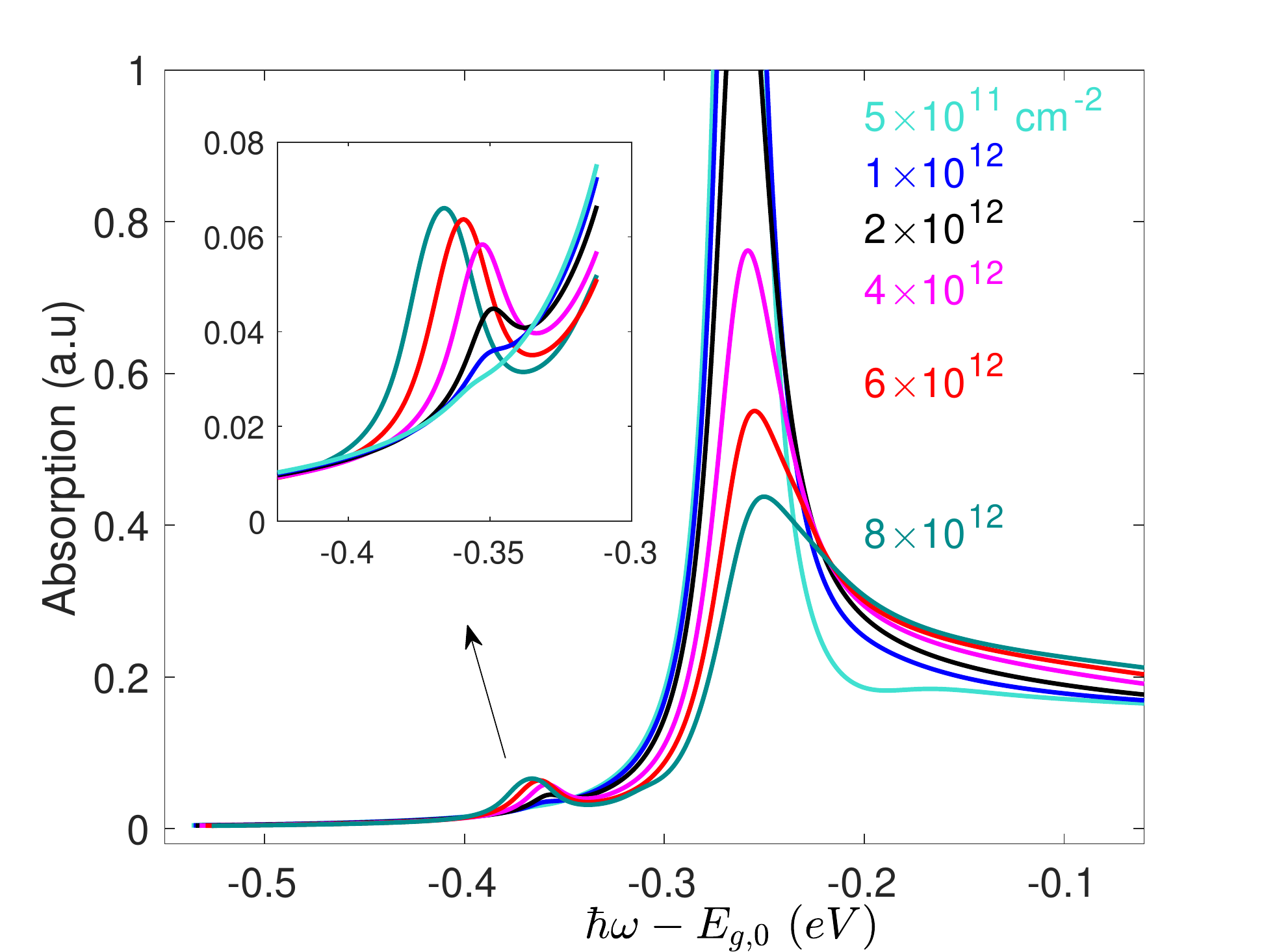}
\caption{  The absorption spectrum of direct excitons in electron-doped ML-WX$_2$ after renormalization due to interaction with shortwave plasmons. A dynamical band emerges in the low-energy side of the absorption spectrum, as highlighted in the enlarged box. The band initially redshifts and intensifies when the electron density increases, showing saturated behavior at elevated densities.}
\label{fig:renorm_WX2}
\end{figure}

Figure~\ref{fig:renorm_WX2} shows the renormalized absorption spectrum of direct excitons in electron-doped ML-WX$_2$ at 10~K. A dynamical band emerges in the low-energy  side of the spectrum. We have considered tungsten-based settings such that indirect excitons have lower energy than direct ones (Fig.~\ref{fig:excitons}(a)). The spin-orbit coupling splits the  conduction-band valleys by 26~meV \cite{Zhang_arXiv17}. All other parameters are the same as in Fig.~\ref{fig:intra}. The behavior of the simulated dynamical band is similar to the empirical behavior of the optical side band in electron-doped ML-WX$_2$ \cite{Jones_NatNano13,Xu_NatPhys14,Shang_ACSNano15,Plechinger_PSS15,Plechinger_NanoLett16,Plechinger_NatCom16,Wang_NanoLett17,Wang_NatNano17}, as shown in Fig.~\ref{fig:exp}(b). Comparing the theory and experiment results, both show a redshift and amplification of the optical side band when increasing the electron density until they saturate.  We note that the pair function reflects the oscillator strength of the optical transition, and in this respect its amplitude should be compared with that seen in differential reflectivity rather than photoluminescence experiments (we will further address this issue later). Comparing the results in Figs.~\ref{fig:exp}(b) and ~\ref{fig:renorm_WX2}, however, we still find two quantitative disagreements. The first one is that the main peak does not completely decay at elevated densities. As explained before, we attribute this disagreement to the fact that broadening effects are introduced only after solving the two-particle Dyson equation (Appendix~\ref{Sec:Computational}).  The second disagreement is that the relative amplitude of the simulated dynamical band is about three times weaker than the oscillator strength of the optical side band. We attribute this mismatch to the difficulty to simulate a stronger dynamical band through increase of the integration cutoff in Eq.~(\ref{eq:self_inter_Dyson}) outside the small window of free-plasmon propagation  (e.g., when $|K_0 - {q}'| \gtrsim  \alpha k_F$). The reason is that the high-order Pad\'{e} polynomials used to perform analytical continuation, $ \Omega \rightarrow  E_{\omega}  + i \delta$, are  notoriously sensitive to minute numerical errors \cite{Osolin_PRB12}.

We now focus on the redshift and initial amplification  of the dynamical band, which are common to both experiment, as shown in Figs.~\ref{fig:exp}(b), and our simulations (highlighted in the enlarged box of Fig.~\ref{fig:renorm_WX2}). The redshift is caused by the electron-electron exchange interaction when the lower spin-split valleys are being filled. This interaction leads to stronger energy redshift of the populated lower valleys compared with the unpopulated upper ones (inset of Fig.~\ref{fig:intra}), thereby increasing the plasmon energy via $\Delta$ in Eq.~(\ref{eq:plasmon_dispersion}). The initial amplification of the dynamical band is commensurate with the increase in the number of free-propagating plasmon modes (larger value of $\alpha k_F/3$). Saturation at high densities occurs when the energy difference  between direct and indirect excitons, $\delta E_\text{D-I}$, deviates from the energy of shortwave plasmons, $\sim \Delta$. Specifically, indirect excitons experience a blueshift due to  electron filling of the lower valleys, while direct ones are not subjected to this effect since they originate from electronic states in the empty upper valleys. Continuing to increase the electron density eventually  leads to a slow decay of the dynamical band due to increased `off-resonance' conditions between $\delta E_\text{D-I}$ and $\Delta$ (Appendix~\ref{Sec:more_results}).  

\begin{figure}[t]
\includegraphics[width=8.5cm]{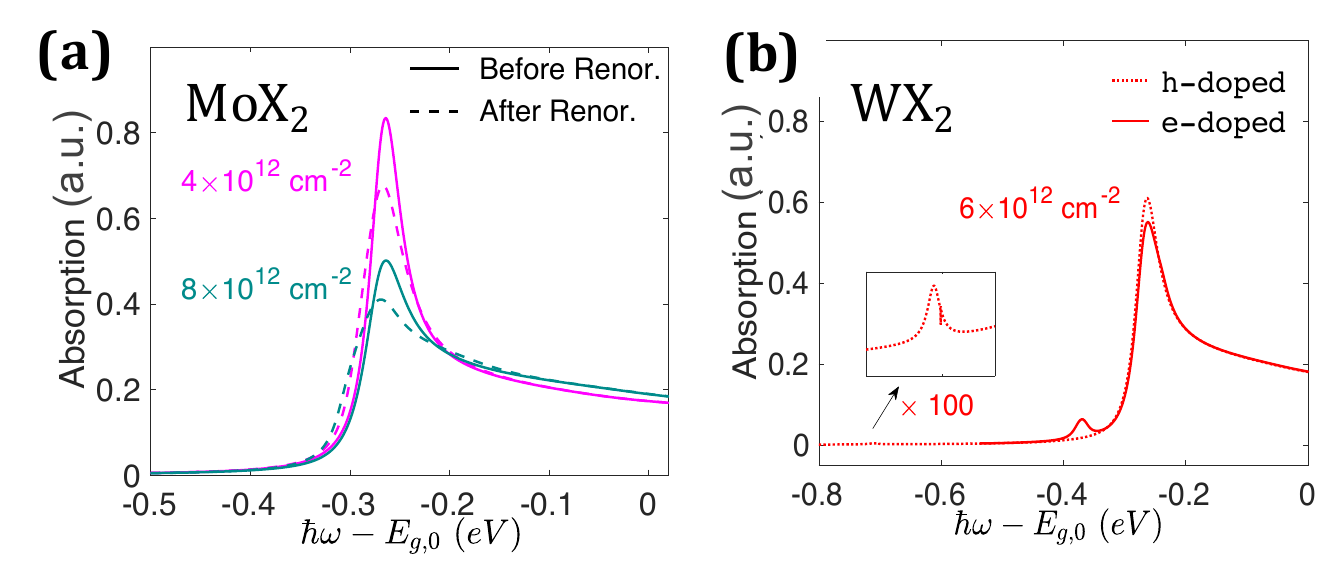}
\caption{  The renormalized absorption spectrum of direct excitons in ML-MoX$_2$ and in the case of hole doping. (a) The case of electron-doped ML-MoX$_2$ before (solid lines) and after (dashed lines) renormalization. The dynamical band is not seen. (b) The case of electron-doped (solid line) and hole-doped  (dotted line) ML-WX$_2$.  The dynamical band is negligible for hole doping, as highlighted in the magnified box. It emerges $\sim$400~meV below the exciton line due to the large valence-band splitting.}
\label{fig:Mo_and_Holes}
\end{figure}

Our model shows that the dynamical band is evident only in electron-doped ML-WX$_2$, providing a strong evidence that it corresponds to the measured optical side band \cite{Jones_NatNano13,Xu_NatPhys14,Shang_ACSNano15,Plechinger_PSS15,Plechinger_NanoLett16,Plechinger_NatCom16,Wang_NanoLett17,Wang_NatNano17}. Figure ~\ref{fig:Mo_and_Holes}(a)  shows that the dynamical band is absent in electron-doped ML-MoX$_2$, simulated by reversing the order of the  conduction-band valleys  such that direct excitons have lower energy (Fig.~\ref{fig:excitons}(b)).  The dynamical band is not seen due to `wrong conditions'. Namely, the band is positioned at about one plasmon energy below the energy of the indirect exciton, which in the case of ML-MoX$_2$ coincides with the direct-exciton spectral line. This behavior makes it hard to resolve the optical signature of exciton-plasmon quasiparticles in ML-MoX$_2$. Figure~\ref{fig:Mo_and_Holes}(a) indeed shows the renormalization only affects the spectral region of the exciton peak without the emergence of dynamical bands in other spectral regions.  In electron-doped ML-WX$_2$, on the other hand, the dynamical band can be distinctly recognized since the indirect exciton has lower energy, so that overall the band emerges $\sim2\Delta$ below the direct-exciton spectral line. 

Further evidence that the revealed exciton-plasmon phenomenon corresponds to the  optical side band in electron-doped ML-WX$_2$ is provided by simulating the hole-doped case. Figure ~\ref{fig:Mo_and_Holes}(b) shows the renormalized absorption spectra of hole- and electron-doped ML-WX$_2$. The dynamical band is practically absent in the hole-doped case, which is similar to the experimental findings in Figs.~\ref{fig:exp}(c) and (d). We can understand this behavior by noting that for electron doping, the splitting of the conduction-band valleys corresponds to both plasmon energy and energy difference between direct and indirect excitons (the role of electron-hole exchange is relatively small as we explain in Appendix~\ref{Sec:eh_exchange}). For hole doping, on the other hand, the plasmon energy is governed by the  large splitting of the valence-band valleys ($\sim$400~meV in ML-WX$_2$), while the energy difference between direct and indirect excitons is still governed by the much smaller splitting in the conduction band. The large mismatch between the two suppresses the emergence of a visible dynamical band in hole-doped ML-TMDs. 


\subsection{Stokes vs anti-Stokes optical side bands} \label{sec:stokes}

Figure~\ref{fig:renorm_WX2} shows that the exciton-plasmon resonance emerges $\sim2\Delta$ below the direct-exciton spectral line in electron-doped ML-WX$_2$. Figure~\ref{fig:Mo_and_Holes}(a) shows that it coincides with the spectral line of the direct exciton in  electron-doped ML-MoX$_2$. These calculations did not reveal the emergence of an optical side band above the direct-exciton spectral line due to the small population of shortwave plasmons at the simulated low temperature (T=10~K), $g(E_{\text{s}}(q)) \rightarrow 0$ when $E_{\text{s}}(q) \sim \Delta \gg k_BT$. Specifically, the self energy of direct excitons includes both plasmon emission and absorption processes, corresponding to `Stokes'  and `anti-Stokes' bands below and above the energy of the indirect exciton, respectively (see Appendix~\ref{Sec:contour}). 
The amplitude of the `Stokes' band is commensurate with  $-g(-E_{\text{s}}(q))=1+g(E_{\text{s}}(q))$, while that of the `anti-Stokes' band with $g(E_{\text{s}}(q))$. The vanishing value of the latter at low temperatures explains the fact that Figs.~\ref{fig:renorm_WX2}~and~\ref{fig:Mo_and_Holes} do not include signatures of high-energy optical side bands due to plasmon absorption. 

At high temperatures, on the other hand, it may be possible to observe a distinct high-energy optical side band in the absorption spectrum of electron-doped ML-MoSe$_2$. The value of $\Delta$ in this material is non-negligible \cite{Dery_PRB15,Kormanyos_2DMater15}, resulting in a well-resolved `anti-Stokes' band whose energy is $\sim2\Delta$ above the spectral line of the direct exciton. This high-energy optical side band is expected to blueshift when increasing the electron density due to the increasing value of $\Delta$. Its amplitude, however, should be measurably weaker than that of the `Stokes' band in ML-WX$_2$ since the ratio $|g(E_{\text{s}}(q))/g(-E_{\text{s}}(q))|$ remains evidently smaller than 1 even close to room-temperature (see Appendix~\ref{Sec:contour}). In addition, the observation of an high-energy optical side band in high-temperature ML-MoSe$_2$  may be elusive if it resides close or within the continuum. That is, the exciton-plasmon coupling is manifested at elevated  electron densities in which the exciton energy approaches the continuum due to the strong BGR. The emergence of an exciton-plasmon resonance above the direct-exciton spectral line is hampered by broadening effects, similar to the behavior of neutral excitons when their energy approaches the continuum (Fig.~\ref{fig:intra}). The plasmon emission process, on the other hand, is relatively robust since the increased value of $\Delta$ with electron density causes the low-energy optical side band  to redshift, thereby keeping it protected from merging into the red-shifting continuum. Accordingly,  broadening effects of the `Stokes' band are mitigated and ultrafast dissociation of the exciton-plasmon quasiparticle to the continuum of  electron-hole pairs is suppressed. Appendix~\ref{Sec:contour} includes further details on the `Stokes' and `anti-Stokes' optical side bands. 


\section{Discussion}\label{sec:only_intra}
When comparing the results in this work with experimental findings, one should recall that the theory only evaluates the pair function and its renormalization due to coupling with plasmons. Accordingly, the theory does not capture optical transitions that are associated with three-body complexes \cite{Suris_PSSB01,Ramon_PRB03,Narvaez_PhysE01,Berkelbach_PRB13,Zhang_NanoLett15,Thilagam_JAP14,Chernikov_PRL14,Ganchev_PRL15,Efimkin_PRB17}, Fermi polarons \cite{Sidler_NatPhys17}, and exciton optical transitions next to localization centers \cite{Hawrylak_PRB91}. Nonetheless, the theory in this work covers two important regimes. The first one is the low density regime, $n \lesssim 10^{12}$~cm$^{-2}$, in which the neutral exciton dominates the absorption spectrum. The dynamical picture we have developed quantifies its induced decay due to long-wavelength plasmons  (Fig.~\ref{fig:intra}), where we find good agreement with reflectivity experiments  \cite{Wang_NanoLett17}. The second case is the high-doping regime, $n \gtrsim 5\times 10^{12}$~cm$^{-2}$, in which both reflectivity and PL experiments show that only one peak survives in the spectrum of electron-doped ML-WX$_2$  \cite{Jones_NatNano13,Wang_NanoLett17}, as shown by Fig.~\ref{fig:exp}(b). The main focus of this work is on this doping regime, and we attribute this optical transition to a new type of quasiparticle where the exciton is coupled to a shortwave charge fluctuation. We provided evidence that the behavior of the optical side band when increasing the electron density matches that of the exciton-plasmon quasiparticle, and further showed that it emerges in electron-doped ML-WX$_2$ but not in hole doped or ML-MoX$_2$ cases (Figs.~\ref{fig:renorm_WX2}~and~\ref{fig:Mo_and_Holes}). Any alternative explanation that aims at deciphering the origin for the optical side band should be consistent with this observation. Below, we discuss several key aspects that one may find helpful when  interpreting experimental findings and when comparing them with our theory.

\subsection{Plasmon-assisted optical transitions} 

We have attributed the optical side band in the reflectance spectra to creation of an exciton-plasmon quasiparticle. It was found by renormalization of the pair function of direct excitons,  through the sum of infinite plasmon-induced virtual transitions to intermediate indirect-exciton states (Fig.~\ref{fig:renorm_Dia}(b)). This physical picture should not be confused with that of a plasmon-assisted optical absorption, in which an indirect exciton is created through two virtual transitions into and from a direct exciton. To better understand this difference and its implication in the context of the experimental data (Fig.~\ref{fig:exp}),  we assume low temperatures where only emission of a shortwave plasmon is feasible during these processes ($\Delta \gg k_BT$). 

\subsubsection{Plasmon-assisted photon absorption (reflectivity spectrum)} 

In photon absorption at low temperatures, the plasmon-assisted process creates an indirect exciton and a shortwave plasmon in the final state. The overall energy conservation of the absorption process is such that the photon energy is the sum of the indirect exciton and plasmon energies,  
\begin{eqnarray}
\hbar \omega &=& E_I(-\mathbf{q}_f) + E_{\text{s}}(\mathbf{q}_f)\,,  \label{eq:abs_plasmon_assisted}
\end{eqnarray}
where the crystal momenta of the plasmon and indirect exciton are opposite to guarantee momentum conservation (the photon carries negligible momentum). Here, the photon is absorbed into a direct-exciton virtual state, followed by plasmon emission that transfers the exciton to its final state. Crystal momentum must be conserved during virtual transitions to and from intermediate states due to translation symmetry. As a second-order process, the absorption amplitude is weak unless the energy of the intermediate virtual state resonates with the real energy level of the direct exciton. Recalling of the energy diagram in Fig.~\ref{fig:excitons}, Eq.~(\ref{eq:abs_plasmon_assisted}) implies that this condition is met in electron-doped ML-WX$_2$  but not in ML-MoX$_2$. That is, the plasmon-assisted photon absorption is a second-order resonance  process in electron-doped ML-WX$_2$, meaning that the neutral-exciton spectral line ($X_0$) has two contributions. The first one is from the first-order absorption process, which creates direct excitons. The second contribution  is from the second-order resonance process, which creates indirect excitons via emission of shortwave plasmons. 

Therefore, an important resulting question is how can one resolve the relative contribution of the plasmon-assisted optical transitions to the oscillator strength of $X_0$ in electron-doped ML-WX$_2$. Clearly, the contribution has negligible weight at vanishing electron densities due to the minute range of free-propagating plasmon modes. When the charge density increases, however, we can identify this contribution by looking for differences between the reflectance data in the electron-doped and hole-doped regimes. Indeed,  Figs.~\ref{fig:exp}(a)  and (c) reveal a clear difference in the measured behavior of $X_0$, whose blueshift and broadening is evidently stronger in the electron doped case. As we suggest below, the plasmon-assisted optical transitions can give rise to this behavior. 

Without the contribution of shortwave plasmons, one would expect the blueshift of $X_0$ to be stronger in the hole-doped case. Specifically, the blueshift of neutral excitons in the absorption spectrum of semiconductors is induced by filling one or both of the bands that are involved in the optical transition \cite{HaugKoch_Book}. This condition is not readily met in electron-doped ML-WX$_2$ since only the bottom conduction  valleys are populated, whereas the optical transition involves the top valleys. Both experiment and theory show that electron  population in the top valleys of ML-WX$_2$ begins at densities that are comparable to  $10^{13}$~cm$^{-2}$ \cite{Dery_PRB15,Wang_NanoLett17}.  The situation is different for the hole-doped case at which hole filling in the top of the valence band is measurable at smaller densities. Thus, while such behavior should lead to weaker or at most comparable blueshift of the direct exciton in the electron-doped case, the experiment shows the opposite. Furthermore, the simulated decay of the pair function in Fig.~\ref{fig:intra} resembles the observed decay of $X_0$ in the hole-doped case better. This fact suggests a missing component in the theory such that dynamical screening and BGR are not enough to fully explain the behavior of $X_0$ in electron-doped ML-WX$_2$. 

Plasmon-assisted optical transitions can complete the puzzle and explain the stronger blueshift and broadening of $X_0$ in electron-doped ML-WX$_2$. There are two complementing facts that support this claim. Firstly, plasmon-assisted optical absorption stems from a resonance  process in electron-doped ML-WX$_2$, suggesting a non-negligible effect. Secondly, Eq.~(\ref{eq:abs_plasmon_assisted}) can explain the  enhanced blueshift in the experiment since the plasmon energy increases at elevated densities (through the increased value of $\Delta$).  Finally, the enhanced broadening of $X_0$ in ML-WX$_2$ when the electron density increases can be explained by closer proximity of the photon energy to the red-shifting continuum (or possibly merging with the continuum). We will provide a complete description of these phenomena alongside second-order perturbation theory to describe the plasmon-assisted optical transitions in a future work.

All in all, the optical side band in the reflectance experiment of electron-doped ML-WX$_2$ ($X-'$) is attributed to the exciton-plasmon quasiparticle, originating from renormalization of the self energy of the direct exciton. Its energy is one (two) plasmon(s) below that of the indirect (direct) exciton. Photon absorption due to plasmon-assisted optical transitions, on the other hand, is a second-order process that converts photons into indirect excitons through plasmon emission at low temperatures. It becomes important in electron-doped ML-WX$_2$, in which  the intermediate virtual exciton state resonates with the real energy level of the direct exciton. The signature of this process in the reflectance experiment of electron-doped ML-WX$_2$ is limited to the spectral region of the direct exciton  ($X_0$). Unlike phonon-assisted optical transitions  \cite{Lax_PR61,Li_PRL10,Li_PRB13,Dery_PRB15}, the amplitude of the plasmon-assisted ones increases with the electron density due to the increased range of free propagating plasmon modes.  

\subsubsection{Plasmon-assisted photon emission (PL  spectrum)} 

In photon emission at low temperatures, the plasmon-assisted process creates a shortwave plasmon and a photon in the final state through  annihilation (radiative recombination) of an indirect exciton. The overall energy conservation of the emission process is such that,  
\begin{eqnarray}
\hbar \omega &=& E_I(-\mathbf{q}_f) - E_{\text{s}}(\mathbf{q}_f)\,.  \label{eq:pl_plasmon_assisted}
\end{eqnarray}
Here, the indirect exciton in the initial state is transitioned into a direct-exciton virtual state by plasmon emission, followed by the photon emission.  Recalling of the energy diagram in Fig.~\ref{fig:excitons}, Eq.~(\ref{eq:pl_plasmon_assisted}) implies that the resonance condition is now met in electron-doped ML-MoX$_2$, in which the energy of the intermediate virtual state resonates with the real energy level of the direct exciton. However, this second-order resonance process has no real value at low temperature PL since the population of indirect excitons in ML-MoX$_2$ is negligible at low temperature: Most excitons in these compounds remain direct after photoexcitation and energy relaxation (Fig.~\ref{fig:excitons}(b)).  Therefore, we do not expect plasmon-assisted optical transitions to have a measurable signature in the PL spectrum of ML-MoX$_2$ at low temperatures.

The case of electron-doped ML-WX$_2$ is unique if we consider the renormalization of direct excitons that led to the exciton-plasmon quasiparticle. In the reflectance spectrum, we have concluded that the contributions from creation of indirect excitons through the plasmon-assisted photon absorption and creation of exciton-plasmon quasiparticles are spectrally resolved: The latter corresponds to the optical side band ($X-'$), and the former can only affect the spectral region of the bare neutral exciton ($X_0$). In the PL spectrum, on the other hand, Eq.~(\ref{eq:pl_plasmon_assisted}) suggests that the photon energy should coincide with that of the optical side band. That is, both the annihilation of the exciton-plasmon quasiparticle and the annihilation of indirect excitons through plasmon-assisted photon emission contribute to $X-'$ in the PL spectrum.  The latter contribution may not be negligible due to large population of indirect  excitons after energy thermalization in ML-WX$_2$ (Fig.~\ref{fig:excitons}(a)). In addition, while the energy of the intermediate virtual state does not resonate with the real energy level of the bare direct exciton, it matches the low energy of the dressed one (i.e., the exciton-plasmon quasiparticle).  Therefore, the recombination of indirect exciton due to plasmon-assisted photon emission may have a measurable signature on the spectral weight of $X-'$, explaining in part why its magnitude is much stronger  in PL compared with reflectance spectra (see related discussion in the Sec.~\ref{sec:intro}) \cite{Jones_NatNano13,Wang_NanoLett17}.  We note that modeling of the density-dependent PL spectrum is beyond the scope of this work, partly because of the lack of knowledge of the non-radiative recombination processes. For example, charged impurities that function as nonradiative recombination centers are screened at elevated densities, resulting in improved quantum efficiency for radiative recombination. Conversely, non-radiative Auger processes are amplified at elevated densities \cite{Hausser_APL90,Polkovnikov_PRB98,Danovich_2DMater16}, thereby reducing the quantum efficiency.  

\subsubsection{High Temperatures} 
So far we have discussed  plasmon-assisted optical transitions at  low temperatures, in which only emission of shortwave plasmons is possible during the photon absorption or emission. At high temperatures, on the other hand, we should also consider the absorption of shortwave plasmons during these optical processes. Repeating the previous analyses, one can see that the conditions for second-order resonance processes are switched between ML-MoX$_2$ and ML-WX$_2$. Detecting the signatures of these processes in the high-temperature reflectance or PL spectra  may be elusive for the same reasons we have mentioned regarding the `anti-Stokes' optical side band in Sec.~\ref{sec:stokes}.

\subsubsection{Other optical spectroscopy tools to probe the plasmon-exciton coupling}

In addition to differential reflectivity and photoluminescence experiments, it may be possible to probe the coupling between excitons and shortwave plasmons in electron-doped samples using Raman-type experiments or non-linear optical spectroscopy. For example, the energy of scattered photons in the `Stokes' band of a Raman experiment is  expected to be 2$\Delta$ below that of the incident photon. It involves a virtual process through which two counter-propagating shortwave plasmons are emitted, so that the sum of their momenta becomes negligible and matches the minute momentum of photons. In the first step of this process, the incident photon induces a virtual transition to a direct exciton intermediate level, which in the second step is scattered to the indirect exciton intermediate level due to plasmon emission. In the next step, the indirect exciton is scattered back to the direct exciton intermediate level by emission of a counter-propagating plasmon. Finally, the system emits a photon at the `Stokes' energy (the energy of the incident photon minus the energy of two plasmons). The overall process is expected to be weak since two plasmon-mediated virtual transitions are involved. The amplitude of the scattered light can increase by using near resonance conditions. The amplitude of the scattered light and its energy can be controlled by employing a gate voltage that tunes the value of $\Delta$ through electron filling of the lower spin-split valleys. That is, the gate voltage provides additional control parameter in the experiment. A similar approach can be used in non-linear wave mixing spectroscopy: The generated wave is amplified when the detuning between the input beams approaches 2$\Delta$. 

\subsection{Three body complexes versus exciton-plasmon quasiparticles}

It is beyond the scope of this work to provide a complete three-body dynamical function since it is impossible to solve numerically its corresponding Dyson equation without simplifying approximations.  The dynamical model we have developed describes the behavior of the pair function and its renormalization due to coupling with shortwave plasmons. The optical side band, which we have associated with the exciton-plasmon quasiparticle, has nearly opposite behavior to that of charged excitons (see discussion of Fig.~\ref{fig:exp} in Sec.~\ref{sec:intro}). Accordingly, the identification of charged excitons complexes and exciton-plasmon quasiparticles can be readily achieved. Studying how the dynamically-screened potential influences the oscillator strength and broadening of  charged excitons is a work in progress. Here, we mention some of the known theoretical approaches to study their behavior. The most computationally convenient way to deal with trions is to dispense with the dynamical terms altogether \cite{Courtade_arXiv17,Narvaez_PhysE01,Berkelbach_PRB13,Zhang_NanoLett15,Thilagam_JAP14,Ganchev_PRL15,Chernikov_PRL14}, but then one can only conclude on their binding energy. Such static models cannot describe the density-dependent rise, decay, energy shift, and broadening of trions in the spectrum. Alternatively, one can neglect dynamical terms in the exciton-electron interaction \cite{Sidler_NatPhys17}, or a priori assume the dynamical form of the Green's function without solving the corresponding dynamical Dyson equation \cite{Suris_PSSB01,Efimkin_PRB17}. In the latter case, one can describe the behavior of charged excitons by invoking an exciton-electron scattering function, where dynamical terms in the scattering potential are neglected.  


One open question in the context of our work deals with possible coupling between trions and shortwave plasmons. Figures~\ref{fig:exp}(a) and (b) do not show optical side bands whose energy is $\Delta$ or $2\Delta$ below that of the charged exciton spectral line; the value of $\Delta$ is about 26~meV at low densities \cite{Zhang_arXiv17}, and it increases at elevated ones \cite{Dery_PRB16}.  The absence or weakness of trion-plasmon coupling is attributed to the relatively small oscillator strength of the charged exciton compared with that of the neutral exciton (Fig.~\ref{fig:exp}). The oscillator strength  is inversely proportional to the square of the complex radius, where the radius is about three times larger for trions \cite{Kylanpaa_PRB15,Mayers_PRB15}. This fact is also reflected in the much smaller binding energy of the charged exciton. Here, we reemphasize that the decay of the neutral-exciton spectral line with the increase in electron density is mostly caused by broadening effects when the exciton energy approaches the redshifting continuum (see discussion of Fig.~\ref{fig:intra}). The quasiparticle made by a neutral exciton coupled to a shortwave plasmon is resilient to this broadening since its energy is kept away from the redshifting continuum due to the increasing value of $2\Delta$ when the density is ramped up. For that reason, the oscillator strength of the optical side band neither decays nor broadens at elevated densities (Fig.~\ref{fig:exp}(b) and Fig.~\ref{fig:renorm_WX2}). Quantitatively, the oscillator strength amplitude is manifested in a relatively large value of the pair function at even Matsubara energies, compared with the respective dynamical Green's function that would describe charged excitons.  Finally, we mention that the optical side band is weak but noticeable even when $E_F$ is  significantly smaller than the binding energy of the charged exciton (enlarged boxes of Figs.~\ref{fig:exp}(a) and \ref{fig:renorm_WX2}). In this regime, the negatively charged exciton manages to stay away from electrons and plasma excitations of the Fermi sea, as evident from the absence of broadening and blueshift of $X^A_-$ and $X^B_-$ in Fig.~\ref{fig:exp}(a). Neutral electron-hole pairs, on the other hand, are not repelled by the surrounding Fermi sea, and therefore can be coupled to shortwave plasmons at all finite densities.  

 \subsection{Strong Photoexcitation conditions}
Recently, You \textit{et al.} observed the emergence of a strong spectral line in the PL of ML-WSe$_2$ when subjected to strong excitation \cite{You_NatPhys15}. This peak showed quadratic dependence on the intensity of the pump signal, and did not shift appreciably when increasing the pumping intensity. This behavior suggests that the emerged spectral line is due to exciton-exciton interaction \cite{You_NatPhys15}.  A similar behavior was also observed in the PL of ML-WS$_2$  \cite{Shang_ACSNano15,Plechinger_PSS15}. To date, however, this behavior was not reported in the PL of ML-MoX$_2$. 

We argue that these observations can be settled by considering the shortwave Coulomb excitations. Specifically, the strong pump intensity generates a relatively large density of direct excitons whose majority relax in energy and turn indirect in ML-WX$_2$ (Fig.~\ref{fig:excitons}(a)) \cite{Dery_PRB15,Zhang_PRL15}. The band structure of indirect excitons is similar to the one shown in Fig.~\ref{fig:plasmons}(a) \cite{Song_PRL13}, where their electron component generates the same type of shortwave Coulomb excitations. A direct exciton can then interact with these excitations in the same way as before leading to renormalization of its self energy. The observed biexciton spectral line emerges through this renormalization. The main difference from the optical side band, which emerges under conditions of weak photoexcitation in electron-doped ML-WX$_2$, is that now the pumping is responsible not only for the exciton population. It is also responsible for the generation of shortwave Coulomb excitations through the electron components of indirect excitons. These two responsibilities explain the quadratic dependence of the observed optical transition on the pumping intensity.   In this view, the measured phenomena can still be identified as an exciton-exciton interaction, but with the notion that the coupling is between direct and indirect excitons due to shortwave charge fluctuations. 

The importance of our theory is therefore in suggesting a microscopic origin for the exciton-exciton optical transition in the PL of strongly photoexcited ML-WX$_2$. We claim that it originates from the intervalley Coulomb excitations, and not due to other sources such as van der Waals-type attraction between two neutral excitons. If the latter was the microscopic origin for the observed biexciton spectral line, then it should be insensitive to whether the two excitons are direct, indirect or a mixture of the two. As a result, one should be able to observe the biexciton spectral line in the PL of both ML-MoX$_2$ and ML-WX$_2$ under comparable pumping intensities. That this scenario is not observed in experiments precludes such microscopic origins. Furthermore, previous theoretical calculations have shown that the binding energy of the biexciton is about 20~meV either in ML-MoX$_2$ or ML-WX$_2$, by using a variety of four-body  computational models with static 2D Coulomb (Keldysh) potentials \cite{Berkelbach_PRB13,Zhang_NanoLett15,Kylanpaa_PRB15,Velizhanin_PRB15,Mayers_PRB15,Kezerashvili_FBS17,Szyniszewski_PRB17}. PL experimental reports, on the other hand,  are available only for ML-WX$_2$ in which the biexciton emerges $\sim\,$50 - 60~meV below the neutral-exciton spectral line \cite{You_NatPhys15,Shang_ACSNano15,Plechinger_PSS15}. Our claim regarding the coupling between direct and indirect excitons through shortwave charge fluctuations is indeed self-consistent with the fact that the biexciton spectral line was not reported to date in PL experiments of ML-MoX$_2$. The low-temperature population of indirect excitons in these compounds is negligible due to their higher energy, as shown in Fig.~\ref{fig:excitons}(b). Moreover, the coupling between indirect and direct excitons has the same `wrong conditions' that were discussed in  the context  of Fig.~\ref{fig:Mo_and_Holes}(a). Namely, the renormalization of direct excitons does not lead to a distinct spectral line, which can be resolved from that of the neutral exciton. 

In addition, our explanation is consistent with the fact that the indirect-direct biexciton spectral line in ML-WX$_2$ emerges at the same spectral position of the optical side band in electron-doped ML-WX$_2$ ($\sim2\Delta$ below the direct-exciton spectral line, which is of the order of $\sim\,$50 - 60~meV in these materials)     \cite{Jones_NatNano13,Shang_ACSNano15,Wang_NanoLett17,Wang_NatNano17}.  One difference between the two cases, however, is that the energy shift of the direct-indirect biexciton is weak when  the photoexcitation intensity increases \cite{You_NatPhys15}, compared with the redshift of the optical side-band when increasing the electron density \cite{Jones_NatNano13}. The reason is that unlike gate-induced free electrons, bound and neutral excitons do not lead to appreciable change of $\Delta$. The value of $\Delta$ can change due to the long-wavelength exchange interaction, which is effective in the case of free electrons  \cite{Dery_PRB16}. As a result, the indirect-direct biexciton shows a relatively weak energy shift when increasing the photoexcitation intensity (as long as excitons remain bound).

\section{conclusion and perspectives}

The identified pairing of shortwave plasmons and excitons explains the puzzling observations of the emerging low-energy optical side band in the spectra of electron-doped WSe$_2$ and WS$_2$, while being conspicuously absent with hole doping, MoSe$_2$ and MoS$_2$ \cite{Jones_NatNano13,Xu_NatPhys14,Shang_ACSNano15,Plechinger_PSS15,Plechinger_NanoLett16,Plechinger_NatCom16,Wang_NanoLett17,Wang_NatNano17}. Elucidating the exciton-plasmon mechanism and the importance of dynamical screening paves the way to harness unexplored manifestations of spin-valley coupling and the realization of other dynamical bands, such  as those arising due to the exciton-phonon coupling. 

The findings of this work reveal the unique nature of shortwave (intervalley) plasmons in  monolayer transition-metal dischalcogenides. Recalling that freely propagating shortwave plasmons  are  independent degrees of freedom in the electronic system, these quasiparticles can be directly detected in electron-rich monolayers (i.e., not through their coupling to excitons).  Using THz spectroscopy,  for example, one should expect to see a resonance in the absorption spectrum when the photon energy is about twice the spin splitting in the conduction band  ($\lesssim$~0.1~eV). Such  far-infrared photons  can couple to two counter-propagating shortwave plasmons, thereby conserving both energy and momentum. As before, the plasmon energy and amplitude can be tuned by a gate voltage. Beyond monolayer transition-metal dichalcogenides, intervalley plasmons may give rise to pairing mechanisms in superconductors whose Fermi surface comprises distinct time-reversal pockets or flat bands. The short-range nature of these plasmons and their gapped energy renders them resilient to screening.

\section*{ACKNOWLEDGMENTS}

The authors are indebted to Zefang Wang, Kin Fai Mak, and Jie Shan for providing the reflectivity measurements in Fig.~\ref{fig:exp} and for many useful discussions. We also thank Xiaodong Xu for discussing and sharing valuable experimental results prior to their publication. This work is mainly supported by the Department of Energy, Basic Energy Sciences (DE-SC0014349). The computational work at Rochester was also supported by the National Science Foundation (DMR-1503601) and the Defense Threat Reduction Agency (HDTRA1-13-1-0013). The work at Buffalo was supported by the Department of Energy, Basic Energy Sciences (DE-SC0004890). The work at Regensburg and W\"{u}rzburg was supported by the DFG (SCHA 1899/2-1 and SFB 1170 ``ToCoTronics''), and by the ENB Graduate School on Topological Insulators.

\appendix

\section{Computational Details}\label{Sec:Computational}

Equation~(\ref{eq:pair_Dyson}) is solved with matrix inversion. Each row (or column) in the matrix is indexed via $\{k_i,z_n\}$. The angular dependence of ${\bf  k}_i-{\bf k}_2$ in $V_s({\bf  k}_i-{\bf k}_2,z-z')$ is averaged out, and then all the 2D wavevectors in Eq.~(\ref{eq:pair_Dyson}) are treated as scalars.  We use a uniform grid with a spacing of $\Delta k=k_{Max}/N_k$ where $k_{Max}$ is chosen such that $\varepsilon_{k_{Max}}=1$~eV. For the Fermion Matsubara energies, $z_n=i\pi (2n_z+1)k_BT$, $n_z$ runs between $-N_z$ to $+N_z$. The Boson Matsubara energy $\Omega$ in Eq.(\ref{eq:pair_Dyson}) is treated as a parameter in the matrix. That is, we solve this equation for each $\Omega_n=i2\pi n_\Omega k_BT$ with $n_\Omega\in[-N_\Omega,N_\Omega]$. For numerical calculations, we have used $N_k=50$, $N_z=160$, $N_\Omega=90$ and the code was written in Fortran  and parallelized with OpenMP. For every density, the calculation of $G_{p}(q=0, k,z,\Omega)$ or $G_{p}(q=K_0, k,z,\Omega)$ takes about 7~hours when using 44 CPUs (2~GHz) with 1~TB shared DRAM. 

After the calculation of the pair Green's function, we use Eq.~(\ref{eq:abs}) to calculate the absorption spectrum. Specifically, we first calculate the function 
\begin{equation}
 A(\Omega) =\beta^{-1}  \sum_{z} \text{Im} \left \{\int dk k  G_p(q = 0, k, z, \Omega)\right \}  \,\,, \label{eq:Aomega}
\end{equation}
and then use Pad\'{e} polynomials to extract the analytical form of $A(\Omega \rightarrow  E_{\omega} + i \delta)$ in the upper complex plane from knowledge of its values at  discrete even Matsubara energies \cite{Vidberg_JLTP77}.  The broadening term is taken by \cite{HaugKoch_Book}
\begin{equation}
 \delta \equiv \delta(E_\omega) = \delta_1 +  \frac{\delta_2}{1 + exp[ (E_c - E_\omega)/\delta_3]}  \,\,, \label{eq:broad}
\end{equation}
where $\delta_j$ ($j=\{1,2,3\}$)  are three parameters with the following meaning: $\delta_1$ reflects the inhomogeneous broadening due to disorder, adatoms, and charge puddles.  $\delta_2$ and $\delta_3$ reflect the enhanced homogenous broadening when the transition energy belongs to the continuum ($E_\omega > E_c$). Far below the continuum, we get that $\delta(E_\omega) \rightarrow \delta_1$.  Far above the continuum edge, we get that $\delta(E_\omega) \rightarrow \delta_1 + \delta_2$. See chapter~13 in Ref.~\cite{HaugKoch_Book} for more details. In this work, we have used   $\delta_1 = 10$~meV, $\delta_2 = 70$~meV, and $\delta_3 = 20$ meV.   

The results shown in Figs.~\ref{fig:renorm_WX2}~and~\ref{fig:Mo_and_Holes} were obtained using the renormalization in Eq.~(\ref{eq:inter_Dyson}). The self-energy term, whose expression is provided in Eq.~(\ref{eq:self_inter_Dyson}), is calculated by summing over even Matsubara energies and by integrating over $\mathbf{q}'$. The integration interval for the latter is $ \left| {\bf q}' -\bf{K_0} \right| \lesssim \gamma \frac{\alpha}{3} k_F $ where $\gamma=2$. That is, we have assumed that plasmons are not immediately damped outside the region of free-plasmon propagation.

\subsection{Effective mass model}\label{Sec:mass}

It is emphasized that the goal of this work is not to reproduce the exact binding energy of excitons in ML-TMDs. Rather, we are interested in the effects of dynamical screening and coupling between excitons and shortwave plasmons. We use the effective mass approximation since it allows us to simplify the calculation while keeping the main dynamical effects intact. In the calculation of the pair functions of direct and indirect excitons, we have assumed similar effective masses for electrons in the lower and upper valleys of the conduction band. We ignore the (small) change in the values of the effective masses at the edges of the lower and upper valleys since the binding energy of excitons in ML-TMDs is much larger than the conduction-valley splitting energy. That is, the matrix we use to solve the pair function includes electronic states whose energies are hundreds of meV above the edge of the band. For such states, the exact value of the effective mass in the bottom of the band is meaningless. We therefore choose the same value for both lower and upper valleys, which represents the effective mass of the `extended` conduction band.  

\subsection{Emission and absorption of plasmons}\label{Sec:contour}

The self energy of the exciton due to the interaction with shortwave plasmons, given in  Eq.~(\ref{eq:self_inter_Dyson}),  includes both emission and absorption terms. Formally, these terms are derived from the identity \cite{Haug_SchmittRink_PQE84,Mahan_Book}
\begin{eqnarray}
&\,&  \beta^{-1} \sum_{\Omega'} D(\Omega - \Omega', \mathbf{q}' )  G_p(\mathbf{q}', \mathbf{k}, \Omega')  \,\,= \\   \label{eq:contour}
&\,& \frac{1}{ 2\pi i}  \oint_C d\Omega' \frac{1}{\exp(\beta\Omega') - 1}  D(\Omega - \Omega', \mathbf{q}' ) G_p(\mathbf{q}', \mathbf{k}, \Omega') \,\,, \nonumber
\end{eqnarray}
where the sum runs over even Matsubara energies, and the integration contour $C$ in the upper plane encircles the poles in the positive sense. Using Cauchy's residue theorem, the Bose Einstein distribution factors for absorption and emission, $g(\pm E_{\text{s}}(q))$, result from the poles in the plasmon propagator, $\Omega' = \Omega \pm E_{\text{s}}(q)$ (see Eq.~(\ref{eq:plasmon_propagator})), where $\exp(\beta\Omega)=\exp(2\pi i  n ) = 1$. Furthermore, by  approximating the pair function as a simple pole expression, $G_p \propto (\Omega' - E_0)^{-1}$ where $E_0$ is the exciton energy, we can analytically derive the 'Stokes' and 'anti-Stokes' side bands. The former stems from the emission term  $\propto g( -E_{\text{s}}(q) )/[\Omega + E_{\text{s}}(q) - E_0]$, where $\Omega \rightarrow \hbar \omega + i \delta$ and $g( -E_{\text{s}}(q) ) \rightarrow 1$ at low temperatures. Similarly, the absorption term stems from $\propto g(E_{\text{s}}(q) )/[\Omega - E_{\text{s}}(q) - E_0] $.  

Section~\ref{sec:stokes} describes some of the difficulties that may arise when one tries to observe the `anti-Stokes' band in the absorption spectrum of high-temperature ML-MoSe$_2$. Other difficulties that impede its detection may be caused by plasmon damping due to coupling with phonons, and by the relatively small population of shortwave plasmons due to the increased value of $E_{\text{s}}(q) \sim \Delta$ at elevated electron densities. This behavior  can be understood from inspecting the Bose-Einstein distribution factors that are associated with the distinct `anti-Stokes' and 'Stokes' bands, $|g(\Delta_{Mo})/g(-\Delta_{W})|$, which emerge in ML-MoSe$_2$ and  ML-WX$_2$, respectively. For example, assuming that these optical side bands persist at 200~K and assigning  $\Delta\sim\,$40~meV for both materials at some comparable elevated densities \cite{Dery_PRB16}, then the amplitude of the 'Stokes'  band becomes $\sim$20 times stronger than that of the `anti-Stokes' band.  


\subsection{Type B excitons}\label{Sec:TypeB}
 Type B excitons correspond to optical transitions from the lower spin-split valence band. We expect the coupling between type B excitons in ML-TMDs and shortwave plasmons to be weak due to the ultrashort lifetime of type B excitons: They experience ultrafast intervalley spin-conserving energy relaxation of  holes to the lower spin-split valleys. Furthermore, at elevated charge densities in which the plasma excitations become relevant, the band-gap renormalization is strong enough so that type B optical transitions are within the continuum. That is, when the densities of resident holes or electrons are larger than $\sim$10$^{12}$~cm$^{-2}$, type B excitons can experience ultrafast dissociation to free electrons and  holes.  The coupling to shortwave plasmons is less effective in this case. 

\subsection{Electron-hole exchange}\label{Sec:eh_exchange}

The relatively large binding energy of excitons in ML-TMDs can lead to evident energy splitting between bright and dark excitons  \cite{Maialle_PRB93,Yu_NatComm14,Echeverry_PRB16}. The theory presented in our work, however, involves only bright excitons (singlet spin configuration of the electron and hole). Therefore, both direct and indirect bright excitons should experience similar effects due to the electron-hole exchange interaction. As a result, the energy splitting between direct and indirect excitons remains equivalent to the spin-splitting of the conduction band. Furthermore, the emergence of the dynamical band is relevant at elevated charge densities in which the wavefunctions of excitons are blown-up due to screening (smaller binding energy). The electron-hole exchange is a weaker effect at such conditions.

\subsection{More simulation results}\label{Sec:more_results}

\begin{figure}[h]
\centering
\includegraphics*[width=9cm]{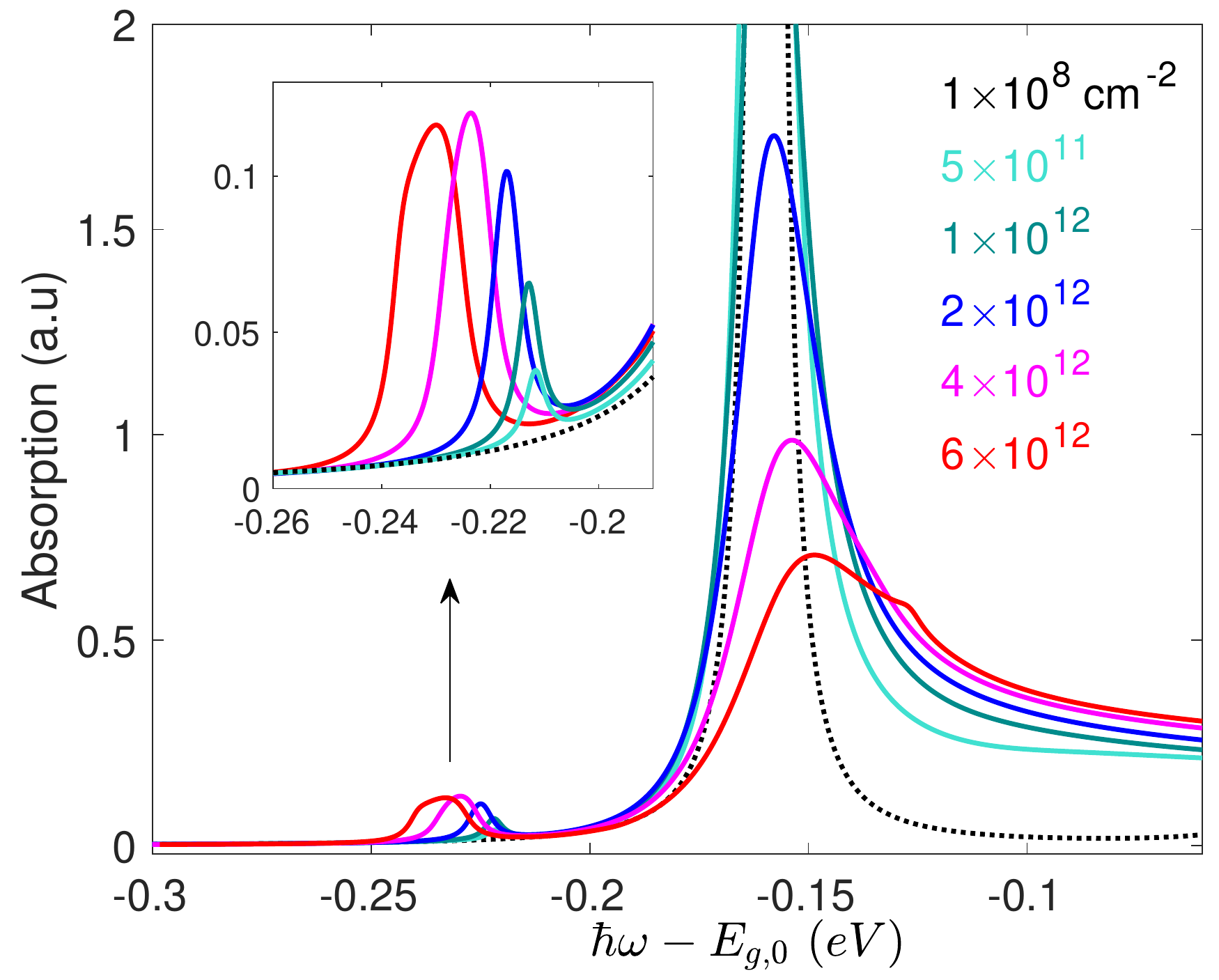}
\caption{ Renormalized absorption spectrum in electron-doped ML-WX$_2$ embedded in layers with dielectric constant, $\epsilon_b = \epsilon_t = 5.06$.  This figure shows the same qualitative behavior as in Figure~4(a) in which we have used  $\epsilon_b = \epsilon_t = 2.7$.}\label{fig:eps506} 
\end{figure}

It is important to realize that the effects of dynamical screening and interaction with plasmons are largely independent of the exact parameters that we choose for the effective mass and dielectric constant. For example, in the main text we have assumed that the monolayer is embedded in hexagonal boron nitride for which there is discrepancy in the reported values of the dielectric constant. Recent experiments use values of  $\epsilon = 2.7$ \cite{Wang_NanoLett17,Wang_NatNano17}, and we have used this parameter in Figs.~\ref{fig:intra}, \ref{fig:renorm_WX2}, and \ref{fig:Mo_and_Holes}. However, previous studies of hexagonal boron nitride reported values that are nearly twice as large, $\epsilon = 5.06$ \cite{Geick_PR66}. We have used this value in Fig.~\ref{fig:eps506}, illustrating that such changes do not  affect the findings of this work.  The only change, as seen from the values of the $x$-axis,  is a smaller binding energy due to the larger  dielectric screening. Yet, the roles of dynamic screening and the emergence  of the dynamical band are the same. Here, we have also used a slightly larger cutoff $ \left| {\bf q}' -\bf{K_0} \right| \lesssim \gamma \frac{\alpha}{3} k_F $ where $\gamma=3$ instead of $\gamma=2$ that we used in Figs.~\ref{fig:renorm_WX2}~and~\ref{fig:Mo_and_Holes}.  We use this value to enable stronger pairing between excitons and shortwave plasmons, as can be seen from the shoulder and enhanced blueshift of the main peak at elevated densities. At the same time, the redshift of the dynamical band is also somewhat enhanced and shows slight decay at elevated densities. 

\end{document}